%% file: baryons_tim_final_edit.tex
\documentclass[fleqn,useAMS,usenatbib]{mn2e}
\usepackage{color}

\usepackage{exscale}
\usepackage{graphicx}
\usepackage{natbib}
\usepackage{rotating}
\usepackage{rotate}
\usepackage{afterpage}
\usepackage{fancyhdr}	
\usepackage{amsmath}
\usepackage{amssymb}
\usepackage{relsize}
\usepackage{lscape}
\usepackage{epsfig}
\usepackage{tabularx}
\usepackage{longtable}
\usepackage{ltxtable}
\usepackage{txfonts}
\usepackage{natbib}
\usepackage{bbm}
\usepackage[outercaption]{sidecap}
\usepackage{afterpage}
\usepackage{widetext}

\input{journaldef.tex}

\newcommand{\ti}{\textit}

\newcommand{\bea}{\begin{eqnarray}}
\newcommand{\be}{\begin{equation}}
\newcommand{\ben}{\begin{enumerate}}
\newcommand{\bi}{\begin{itemize}}
\newcommand{\eea}{\end{eqnarray}}
\newcommand{\ee}{\end{equation}}
\newcommand{\ei}{\end{itemize}}
\newcommand{\een}{\end{enumerate}}
\newcommand{\nn}{\nonumber}

\newcommand{\matC}{\mathbf C}

\newcommand{\like}{L}
\newcommand{\prob}{P}
\newcommand{\probr}{P_r}
\newcommand{\pco}{\vek p_\mr{co}}
\newcommand{\pnu}{\vek p_\mr{nu}}
\newcommand{\D}{\vek D}
\newcommand{\Del}{\vek \Delta}
\newcommand{\M}{\vek M}

\newcommand{\U}{\vek U}

\newcommand{\om}{\Omega_\mr m}
\newcommand{\omb}{\Omega_\mr b}
\newcommand{\sig}{\sigma_8}
\newcommand{\ns}{n_s}
\newcommand{\w}{w_0}
\newcommand{\wa}{w_a}

\renewcommand{\d}{{\rm d}}
\newcommand{\pd}{P_{\delta}}

\newcommand{\eps}{\epsilon}
\newcommand{\abs}[1]{| #1 |}
\newcommand{\mr}{\mathrm}

\renewcommand{\d}{{\rm d}}

\def\vek{\mathbf}

\title[Accounting for baryonic effects on cosmic shear tomography]{Accounting for baryonic effects in cosmic shear tomography: Determining a minimal set of nuisance parameters using PCA}
\author[Eifler et al.]
{\parbox{\textwidth}{Tim Eifler$^{1,2}$,\thanks{E-mail: \texttt{tim.eifler@jpl.nasa.gov}}
Elisabeth Krause$^{1}$,
Scott Dodelson$^{3,4,5}$,
Andrew R. Zentner$^{6,7}$,
Andrew P. Hearin$^{5,8}$,
Nickolay Y. Gnedin$^{3,4,5}$\vspace{0.4cm}}\\
\parbox{\textwidth}{$^{1}$ Department of Physics and Astronomy, University of Pennsylvania, Philadelphia, PA 19104, USA \\
$^{2}$ Jet Propulsion Laboratory, California Institute of Technology, 4800 Oak Grove Dr., Pasadena, CA 91109\\
$^{3}$ Kavli Institute for Cosmological Physics,  Enrico Fermi Institute, University of Chicago, Chicago, IL 60637, USA\\
$^{4}$ Department of Astronomy \& Astrophysics, University of Chicago, Chicago, IL 60637, USA\\
$^{5}$ Fermilab Center for Particle Astrophysics, Fermi National Accelerator Laboratory, Batavia, IL 60510-0500, USA\\
$^{6}$ Department of Physics and Astronomy, University of Pittsburgh, Pittsburgh, PA 15260, USA\\
$^{7}$ PITTsburgh Particle physics, Astrophysics, and Cosmology Center (PITT PACC), University of Pittsburgh, Pittsburgh, PA 15260, USA\\
$^{8}$Yale Center for Astronomy \& Astrophysics, Yale University, New Haven, CT\\
}}

\begin{document}

\date{accepted received}

\maketitle

\label{firstpage}

\begin{abstract}
Systematic uncertainties that have been subdominant in past large-scale structure (LSS) surveys are likely to exceed statistical uncertainties of current and future LSS data sets,  potentially limiting the extraction of cosmological information. 
Here we present a general framework (\ti{PCA marginalization}) to consistently incorporate systematic effects into a likelihood analysis. 
This technique naturally accounts for degeneracies between nuisance parameters and can substantially reduce the dimension of the parameter space that needs to be sampled.\\
As a practical application, we apply PCA marginalization to account for baryonic physics as an uncertainty in cosmic shear tomography. Specifically, we use \textsc{CosmoLike} to run simulated likelihood analyses on three independent sets of numerical simulations, each covering a wide range of baryonic scenarios differing in cooling, star formation, and feedback mechanisms. We simulate a Stage III (Dark Energy Survey) and Stage IV (Large Synoptic Survey Telescope/Euclid) survey and find a substantial bias in cosmological constraints if baryonic physics is not accounted for. We then show that PCA marginalization (employing at most 3 to 4 nuisance parameters) removes this bias.
Our study demonstrates that it is possible to obtain robust, precise constraints on the dark energy equation of state even in the presence of large levels of systematic uncertainty in astrophysical processes. We conclude that the PCA marginalization technique is a powerful, general tool for addressing many of the challenges facing the precision cosmology program.
\end{abstract}


\section{Introduction}
\label{sec:intro}

The increased quality and size of data sets from ongoing wide-field imaging surveys, such as Kilo-Degree Survey (KiDS\footnote{http://www.astro-wise.org/projects/KIDS/}), Hyper Suprime Cam (HSC\footnote{http://www.naoj.org/Projects/HSC/HSCProject.html}), and Dark Energy Survey (DES\footnote{www.darkenergysurvey.org/}), will shift the focus of cosmological analyses from the statistical precision with which a signal is measured to the robustness of the cosmological constraints that are derived from the measurements. Our ability to understand, constrain and model systematics will play a key role in removing biases and reducing the error bars on cosmological parameters; this will be even more crucial for the success of future ground- and space-based endeavors such as the Large Synoptic Survey Telescope (LSST\footnote{http://www.lsst.org/lsst}), Euclid\footnote{sci.esa.int/euclid/} and the Wide-Field Infrared Survey Telescope (WFIRST\footnote{http://wfirst.gsfc.nasa.gov/}).

Cosmological analyses of imaging surveys are affected by a variety of systematic uncertainties. The most important systematics for contemporary and next generation (Stage III and IV according to \cite{abc06,wme13}) surveys are photometric redshift errors, shear calibration, galaxy bias, baryonic physics, intrinsic alignments, and modeling the non-linear evolution of the density field. Uncertainties from these sources are generally expressed through so-called \textit{nuisance parameters} over which one \textit{marginalizes} in a likelihood analysis. The term \textit{nuisance parameter} refers to any parameter in a likelihood analysis except those one aims to constrain. Many of the aforementioned sources of systematics are  interesting astrophysical phenomena in and of themselves, and constraining these phenomena will henceforth go hand in hand with any successful cosmological analysis.

In the literature, the topic of nuisance parameters has been covered extensively. Most of the work to date has considered one or at most two particular systematics, outlining methods to incorporate them into a likelihood analysis. Prominent examples are \cite{mhh06}, \cite{beh10}, \cite{hzm10} for photo-z uncertainty, \cite{hir03} or \cite{hkm05} for shear calibration, \cite{his04} or \cite{jma11} for intrinsic alignment, \cite{jzl06}, \cite{zrh08}, \cite{shs11}, \cite{zsd13}, and \cite{shs13} for the impact of baryonic physics (a topic of immediate interest for the present paper), \cite{zwb11}, \cite{clb12}, \cite{khm13}, \cite{zhb14}, and \cite{rtw14} for galaxy bias/Halo Occupation Distribution modeling. This list is far from complete; defining and constraining nuisance parameters is an active research topic. 

Some of these parameterizations are physically motivated and address specific effects (e.g., halo concentration for baryons, redshift scaling and power spectrum amplitude for intrinsic alignment, multiplicative and additive shear bias, etc). In the absence of information on the functional form of the nuisance parameterization one must rely on introducing distinct nuisance parameters in bins of redshift and scale \citep{ber09,job10}, so as to absorb a variety of possible systematic errors, and rely on the data to calibrate these nuisance parameters. When carrying out a combined probes analysis \citep[as in][for example]{eks14}, where not one but all of these nuisance parameters must be considered simultaneously, the shear number of nuisance parameters challenges the limit of computationally feasibility.

In this paper, we develop a \textit{ Principal Component Analysis (PCA) marginalization} framework that poses an efficient method to incorporate many nuisance parameters and many systematic errors within a likelihood analysis. This framework identifies the principal components (PCs) that capture the impact of nuisance parameters on the quantity that enters the likelihood analysis (e.g., power spectra, correlation functions, etc.). The marginalization procedure can then be carried out efficiently in the PC basis.

We apply this framework to a specific example, namely the impact of various baryonic scenarios on cosmological constraints from weak lensing tomography. Weak lensing tomography is one of the core cosmological probes of photometric surveys; independent of any assumptions about the relationship between dark and luminous matter, weak lensing tomography provides valuable information about the geometry and structure growth of the Universe and thereby allows us to constrain cosmology \citep{hyg02,wmh05,jjb06,shj10,lds12,hwm12,heh14}. In combination with accurate redshift information, weak lensing tomography has been identified as one of the most powerful tools to constrain the dark energy equation of state and thereby reveal the nature of the acceleration of the expansion of the Universe \citep{abc06,pse06,wme13}.

A potentially significant source of systematic error for weak lensing tomography is theoretical uncertainty in the role of baryonic physics in our Universe. Baryonic processes can redistribute matter within the Universe to a degree that is large enough to induce significant systematic errors in cosmological parameters \citep{zrh08,hez09,shs11,shs13,zsd13}, yet the baryonic processes that drive galaxy formation and evolution remain poorly understood and poorly constrained. Different treatments of baryonic gas cooling, star formation, and feedback mechanisms can dramatically alter the predictions for shear measurements (especially on small angular scales), and this effect introduces an intolerable bias in the cosmological parameter estimation. 

In this paper, we examine different baryonic scenarios from 3 independent hydrodynamical simulation efforts: The OWLS (OverWhelmingly Large Simulations) project \citep{sdb10,dsb11}, the simulations used in \cite{rzk08}, and a yet unpublished set of Hydro simulations further described in Sect. \ref{sec:sim}. We simulate a DES and LSST/Euclid likelihood analysis in a 7-dimensional cosmological parameter space using the PCA marginalization scheme to take baryonic uncertainties into account.

\section{Marginalization of Baryonic Effects}

\subsection{Likelihood Analysis Basics}

Given a data vector $\D$ we calculate the posterior probability for a point in the joint parameter space of cosmological parameters $\pco$ and nuisance parameters $\pnu$ via Bayes' theorem
\be
\label{eq:bayes}
\prob(\pco, \pnu|\D) \propto \probr (\pco, \pnu) \,\like (\D| \pco, \pnu),
\ee
where $\probr (\pco, \pnu)$ denotes the prior probability and $\like (\D| \pco, \pnu)$ is the likelihood. The data vector includes, for example, two-point functions in the form of power spectra, which depend on both scale and redshift. 
The likelihood is often assumed to be Gaussian so that
\be
\label{eq:like}
\like (\D| \pco, \pnu) = N \, \times \, \exp \biggl( -\frac{1}{2} \underbrace{\left[ (\D -\M)^t \, \matC^{-1} \, (\D-\M) \right]}_{\chi^2(\pco, \pnu)}  \biggr) \,.
\ee
We abbreviate $\M=\M(\pco, \pnu)$, i.e. the model vector $\M$ is a function of cosmology and nuisance parameters. The normalization constant $N=(2 \pi)^{-\frac{n}{2}} |C|^{-\frac{1}{2}}$ in Eq. (\ref{eq:like}) can be neglected under the assumption that the covariance is constant in parameter space.
We note that assuming a constant, known covariance matrix $\matC$ is an approximation to the correct approach of a cosmology dependent or estimated covariance \citep[see][for further details]{esh09}. The impact of this assumption on cosmological constraints is more severe for deep, small surveys and less important for wide, shallow surveys.

\subsection{Mode Removal - PCA marginalization}
\label{sec:likebasics}

Consider an experiment that provides a data vector $\D$, which in our case is the set of all auto- and cross-spectra $C_l^{ij}$ of cosmic shear across redshift bins with indices $i,j$. For any set of cosmological parameters, dissipationless N-body simulations are sufficient to produce an accurate prediction for this data vector if dark matter alone were responsible for the lensing. Let us call this prediction, $\M_0(\pco)$, where again $\M$ is a vector with all auto- and cross-spectra and the subscript denotes that the prediction is generated assuming the Universe contained only dark matter and no baryons; the prediction, of course, {\em would} depend on the set of cosmological parameters $\pco$. 

The true prediction for this set of cosmological parameters including the effects of baryons is far more challenging to make. In principle, such a prediction involves a whole new suite of parameters, $\pnu$, that encode the effects of baryons on large-scale structure. If it were possible to specify those parameters and easily generate a prediction for weak lensing power spectra for each parameter set, then we could calculate the likelihood function for the cosmological parameters by marginalizing over the nuisance parameters:
\be
\label{eq:likemarg}
L (\D|\pco ) = \int \d \vek{ \pnu}  \, \exp \biggl( -\frac{1}{2} 
(\D - \M(\pco,\pnu))^t \matC^{-1} (\D - \M(\pco,\pnu)) 
\biggr) \,,
\ee
where $\matC$ is the covariance matrix (which we approximate to be independent of any of the parameters).
Several groups have tried to implement this idea, most successfully by parametrizing baryonic effects with several {\it halo model} parameters \citep{zrh08,shs11,zsd13,shs13}. 

We introduce an alternative way to carry out the marginalization, which does not require detailed understanding of the underlying phenomenology, nor an analytical model associated with the parameters encoding the effects. Rather, this marginalization is over the linear combinations of observables that are most strongly influenced by the baryonic effects (or, more generally, by the systematic of interest). If these modes can be identified, they can easily be integrated out. So, even without any explicit parametrization of the underlying physics, one can account for the associated systematic effects. 

To identify the offending modes, we start with a suite of hydrodynamic simulations, each of which generates a prediction $\M_\alpha(\pco)$. The subscript $\alpha$ refers to the considered numerical simulation and ranges up to the total number of baryonic scenarios $N_\mr{sce}$, which is of order 15 in in our analysis. There is also a dark matter only simulation which, as mentioned above, is identified by $\alpha=0$. The components of the \ti{difference matrix} $\vek \Delta$ between the hydrodynamical simulations and the dark matter only simulation are obtained as
\be
\label{eq:diffmat}
\Delta_{k\alpha} \equiv M_{k\alpha} - M_{k0},
\ee
where the index $k$ here covers all $l$ for all auto- and cross-spectra (that is, $k$ runs over all observables). The difference between the parametric and non-parametric approach is beginning to emerge. In the parametric approach, $\Delta_{k\alpha}$ would be a function of the nuisance parameters; here it is simply a number that captures the uncertainties due to baryonic effects. 

Before proceeding to the general procedure we propose here, consider first a trivial, but instructive example. Suppose that all the hydro simulations predict that all the spectra are identical to the DM-only spectrum except at a single value of $C_l^{ij}$, so that $\Delta_{k\alpha} = 0$ for all $\alpha$ and all $k$ except for $k=1$ (so $\Delta_{k\alpha} \propto \delta_{1\alpha}$ and this first observable corresponds to, say $C_{l=100}^{11}$). A very simple way to deal with the systematic would simply be to remove that single measurement. This is equivalent to setting $\M(\pco,\pnu) = \M_0(\pco) + \delta_{1\alpha} A$, where $A$ is an arbitrary amplitude, and integrating over all possible values of $A$. 

In other words, instead of integrating over parameters $\pnu$, we are integrating over amplitudes of offending modes, where a {\it mode} is a linear combination of all the $C_l^{ij}$ (all the observables). In this simple example, there is only one mode and the coefficients in the linear combination that define that mode are all zero except for one. More generally, a given mode will depend on all the elements of the auto- and cross-spectra, and there could be more than one mode that is marginalized over.

The only remaining difficulty is to identify the modes that are most damaging. There are several ways to approach this. Here we choose to remove modes that have the largest variance in the simulations. To identify the modes with the largest variance, we collect the $\vek \Delta_\alpha$'s from all the simulations into a single matrix $\vek \Delta$. To be concrete, we consider 14 (for DES) and 12 (for LSST/Euclid) simulations so $\vek \Delta$ has 14 (12) columns. We assume five redshift bins so that the total number of auto- and cross-spectra is $5*6/2=15$. We bin so that each spectrum is sampled at 20 values of $l$, meaning that there are a total of 300 data points. So the matrix $\vek\Delta$ has 14 (12 for LSST/Euclid) columns and 300 rows.

The matrix product $\vek\Delta \vek\Delta^t$ is proportional to the covariance of the observables among all of the different baryonic simulations (the $M_{k\alpha}$) with respect to the DM-only simulation ($M_{k0}$). 
Identifying the linear combinations of observables most susceptible to contamination from baryonic processes amounts to diagonalizing the matrix $\vek \Delta \vek \Delta^t$ and choosing the eigenvectors (which are linear combinations of observables) with the largest eigenvalues (the largest variances). The matrix we aim to diagonalize is the product $\vek \Delta \vek \Delta^t$ and we will need to project observables onto the eigenvectors of this matrix, so it is convenient to proceed using the (full) singular value decomposition (SVD) of $\vek\Delta$, 
\be 
\label{eq:SVD}
\vek \Delta = \U \mathbf \Sigma \mathbf V^t \, .
\ee
The PCs of $\Del$ are the columns of the orthogonal matrix $\U$, which in our example is $300\times300$. The mean squared deviations of the observables from the DM-only predictions are 
 \be
 \label{eq:singtocov}
 \mr{Cov} \Del = \frac{1}{N_\mr{sce}-1} \Del \, \Del^t  = \frac{1}{N_\mr{sce}-1} \U \, \mathbf \Sigma \, \mathbf \Sigma^t \, \U^t  =\U \, \mathbf E \, \U^t\,,
 \ee
 where $\vek E =  \frac{1}{N_\mr{sce}-1} \mathbf{\Sigma}\mathbf{\Sigma}^t$ is a diagonal matrix whose (300) entries are the eigenvalues of $\mr{Cov}\Del$. 

We can project the observables onto the PCs in $\U$. We can then identify the linear combinations (or ``modes") most susceptible to baryonic effects as those with the largest entries $E_k$ in $\mathbf{E}$ and remove them from the analysis (equivalent to marginalizing over a free amplitude for them).  In this way, we simply discard the information contained within these modes just as we discarded the information in the observable $k=1$ ($C_{\ell=100}^{11}$) in our pedagogical example above.

Proceeding further requires a bit of care, because the mode must be removed from both the data and the model, so we explicitly walk through our algorithm. 
At each point in cosmology sampled by the MCMC we compute the matrix $\Del$ and obtain the corresponding projection matrix $\vek U^t$ via SVD  as in 
Eq.~(\ref{eq:SVD}). Since $\U$ is an orthogonal matrix, which implies $\mathbbm{1}=\U \U^t = \U^t \U$, we rewrite $\chi^2(\pco, \pnu)$ as 
\be
\label{eq:chipc}
\chi^2(\pco, \pnu) =  (\D - \M )^t \U \U^t \matC^{-1} \U \U^t  (\D - \M )  
\ee
We can then insert a projection matrix $\vek P=\vek P^2$ into Eq. (\ref{eq:chipc}) to restrict attention to a subset of the observables, yielding a new 
$\chi^2(\pco,\pnu)$,  
\be
\label{eq:chipc2}
\chi'^2(\pco, \pnu) = (\vek P \U^t \D - \vek P \U^t \M )^t (\vek P \U^t \matC \U \vek P)^{-1}   (\vek P \U^t \D - \vek P \U^t \M)\,.  
\ee
If $\vek P = \mathbbm{1}$ we recover $\chi'^2(\pco,\pnu) = \chi^2(\pco, \pnu)$ as defined in Eq. (\ref{eq:like}); 
setting some of the diagonal elements in $\vek P$ to zero projects onto a subspace of the PCs. Below, we experiment with the number of modes that need to be removed such 
that the nuisance parameters need no longer be accounted for explicitly in the model: we will see that very few are needed in order to eliminate the systematic of baryonic effects.

Before showing our results, we make two general remarks. First, our choice of which modes to remove is not necessarily the optimal choice. Another well-motivated choice would result if one were to weight the covariance in Eq. (\ref{eq:singtocov}) with the inverse of the data covariance matrix $\matC$. Returning to our simple example of a single observable ($C_{\ell=100}^{11}$) comprising a mode, if the noise in a particular survey at that mode were very large, it would not make sense (or be necessary) to remove the mode. That is, the large variation of a mode alone does not guarantee that it will produce parameter bias. If the mode is not well measured, it is not necessary to remove the mode. Yet another example would be to choose to remove modes that most affect the inferred cosmological parameters of interest. Some modes may exhibit little degeneracy with the parameters of interest and consequently, removing those modes should be a lower priority. 

The second comment is that, while we focus here on the systematic of baryonic effects on the lensing spectrum, the PCA marginalization approach can be applied generally to any probe and any number of systematics. We will address both issues in Sect. \ref{sec:general}.

\section{Uncertainties in baryonic physics}
\label{sec:baryons}

In the following we consider the uncertainties in modeling baryonic physics in weak lensing. We examine various baryonic scenarios from different sets of simulations and calculate shear tomography power spectra for each scenario considering a DES and a LSST/Euclid  like survey (see Table \ref{tab:survey} for details). The values for DES stem from DES documents and internal communication within the DES collaboration; for LSST/Euclid we rely on specifications outlined in \cite{cjj13}. Although \cite{cjj13} aims at LSST only, Euclid survey parameters are similar \citep[15000 deg$^2$, 30 $n_\mr{gal}$, according to][]{laa11}.

The main difference between Euclid and LSST (aside from observational systematics) is the redshift distribution of source galaxies, where Euclid is shallower compared to LSST. It is however unlikely that this difference qualitatively affects the outcome of the analysis presented here, hence we believe that the LSST scenario very well resembles the Euclid survey as well.

\begin{table}
\caption{Survey parameters}
\begin{center}
\begin{tabular}{|l|c c c c c c|}
\hline
\hline
Survey & area $[\mr{deg^2}]$ & $\sigma_\epsilon$ &  $n_\mr{gal}$ &$z_\mr{max}$ & $z_\mr{mean}$ & $z_\mr{med}$ \\
\hline
DES & $5,000 $ & $0.26$  & $10$ & 2.0 & 0.84 &0.63\\
LSST/Euclid & $15,000$ & $0.26$ &$31$ & 3.5 & 1.37 & 0.93 \\
\hline
\hline
\end{tabular}
\end{center}
\label{tab:survey}
\end{table}

\subsection{Simulation Set}
\label{sec:sim}
\begin{table}
\caption{Summary of the baryonic physics in the OWLS simulations.}
\label{tab:1}
\begin{tabular}{l l}\hline
Simulation  &Description   \\ \hline
DM   & No Baryons, CDM only  \\
REF  &  Chabrier (2003) IMF, Wind mass loading $\eta=2$, \\
& $v_\mr{w}=600\, \mr{km\, s^{-1}} $\\
AGN   & Includes AGN (in addition to SN feedback)  \\
NOSN  & No SN energy feedback  \\
$\mr{NOSN\_NOZCOOL}$ & No SN energy feedback and   \\
& cooling assumes primordial abundance\\
 NOZCOOL  & Cooling assumes primordial abundance \\
WDENS  & Wind mass loading and velocity depend on  \\
& gas density (SN energy as REF)  \\
WML1V848   & Wind mass loading $\eta=1$, velocity  \\
& $v_\mr{w}=848\, \mr{km\, s^{-1}} $ (SN energy as REF)\\
WML4   &  Wind mass loading $\eta=4$ (SN energy as REF) \\
DBLIMFV1618 & Top-heavy IMF at high pressure, \\
& extra SN energy in wind velocity \\
\end{tabular}
\end{table}
\paragraph*{OWLS simulations} From the OWLS project we obtain matter power spectra for nine different scenarios corresponding to various hydrodynamical recipes that differ in their treatment of cooling, SN- and AGN feedback. Please see Table \ref{tab:1} for a brief summary and \cite{sdb10}, and \cite{dsb11} for a detailed description of the implemented physics and the observations that motivated these recipes. 
The OWLS simulations were conducted in cubic simulation volumes with sides of length $L=100\, h^{-1}\mathrm{Mpc}$ and the  simulation power spectra have been tabulated by 
\cite{dsb11} and are valid for wave numbers $0.314 \le k/h\mathrm{Mpc}^{-1} \le 10$. These simulations were analyzed for a similar application using a different technique in \cite{zsd13} and \cite{shs13}.\\   

\paragraph*{Rudd simulations} 
The simulations of \cite{rzk08} track the formation of structure in a cubic volume $60\,h^{-1}\mathrm{Mpc}$ on a side in a flat, $\Lambda$CDM cosmological model with $\Omega_{\mathrm{M}}=0.3$, $\Omega_{\mathrm{B}}h^2=0.021$, $h=0.7$, and $\sigma_8 = 0.9$. The simulation set consists of three simulations all starting from the same initial conditions. The first simulation \citep[labeled ``DMO" in][]{rzk08} is purely dissipationless and includes a collisionless dark matter component only. The second simulation (labeled ``DMO\_NR") follows both dark matter and baryons. However, the baryonic component is not permitted to cool radiatively in DMG\_NR. The baryonic component in DMG\_NR is treated in the non-radiative (or ``adiabatic") regime and neither stars nor galaxies form in DMG\_NR. The third simulation (labeled ``DMG\_SF") treats the baryonic component including radiative cooling and heating, star formation, and feedback from supernovae. The inclusion of these processes in DMG\_SF allows for the formation of galaxies in the DMG\_SF simulation. The cool gas forms a condensed component, a fraction of which is converted into stars according to a relatively standard, observationally-motivated star formation recipe.

The dissipationless DMO simulation is performed using the Adaptive Refinement Tree (ART) N-body code \citep{kkk97,kra99}. In the DMG\_NR and DMG\_SF simulations, the gaseous baryonic component is simulated using using an Eulerian hydrodynamics solver on the same adaptive mesh of the N-body ART code using the techniques described by \cite{kkh02}. However, the two simulations that included baryons are performed with the new, distributed-memory version of the N-body+gas dynamics ART code.

A common problem in studies of this kind is that simulations that resolve galaxy formation necessarily model fairly small volumes. The Rudd et al. simulations are among the smaller simulations (computational cube with a side length of $60\,h^{-1}\mathrm{Mpc}$) used for these purposes. Consequently, cosmic variance and finite volume effects are significant at scales of $k \lesssim 0.11\, h\mathrm{Mpc}^{-1}$.

\paragraph*{Gnedin simulations\footnote{publicly available at http://astro.uchicago.edu/$\sim$gnedin/WL/}} 

Four new sets of simulations are performed with the Adaptive Refinement Tree (ART) code (the same code used for Rudd simulations). Each set includes 3 different random realizations with different values for the DC mode \citep{gkr11} of a 200$h^{-1}$ comoving Mpc box with $512^3$ dark matter particles and a factor of several larger number of adaptively refined cells (which are dynamically created and destroyed in the course of the simulation to maintain required spatial resolution). Spatial resolution (the size of the most refined cells) of all simulations is set to $3h^{-1}$ comoving kpc. The first set of simulations is dissipationless and treats dark matter only. The second set (AD) includes only "adiabatic" (i.e.\ non-radiative) hydrodynamic processes. The third set (CW) includes radiative cooling (but no radiative heating) with primordial abundances of hydrogen and helium. The fourth set (CX) includes radiative cooling with the cooling function that corresponds to solar-metallicity gas; that cooling function is applied to all gas in the simulation, even to the deepest voids, and, hence, is physically unrealistic. The CX set should, therefore, be considered as an extreme limit of gas cooling.

\begin{figure}
  \includegraphics[width=\columnwidth]{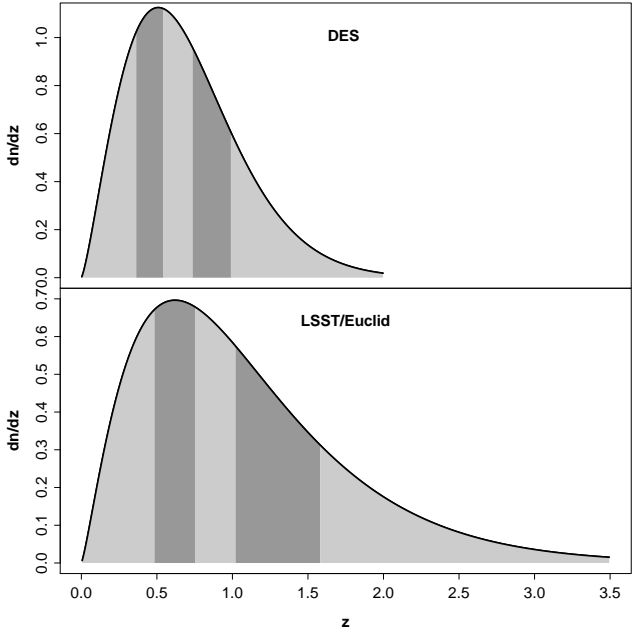}
\caption{The assumed redshift distribution with 5 tomography bins for DES (\ti{top}) and LSST/Euclid (\ti{bottom}).}
 \label{fi:zdistrib}
\end{figure}

\begin{figure}
    \includegraphics[width=8.5cm]{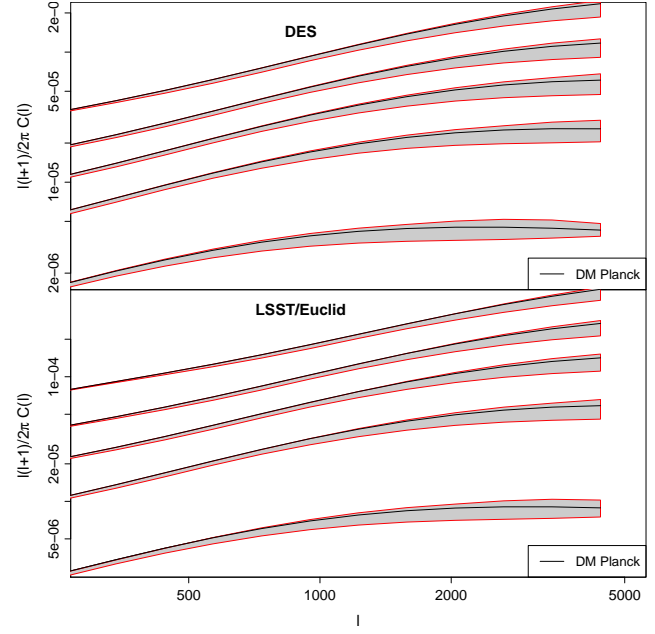}
    \caption{The shear tomography power spectra for the five auto z-bins computed at the fiducial cosmological model. The black line corresponds to the dark matter scenario, the shaded area spans the range of uncertainty from baryonic physics. } \label{fi:cl_final}
\end{figure}

\begin{figure}
    \includegraphics[width=8.5cm]{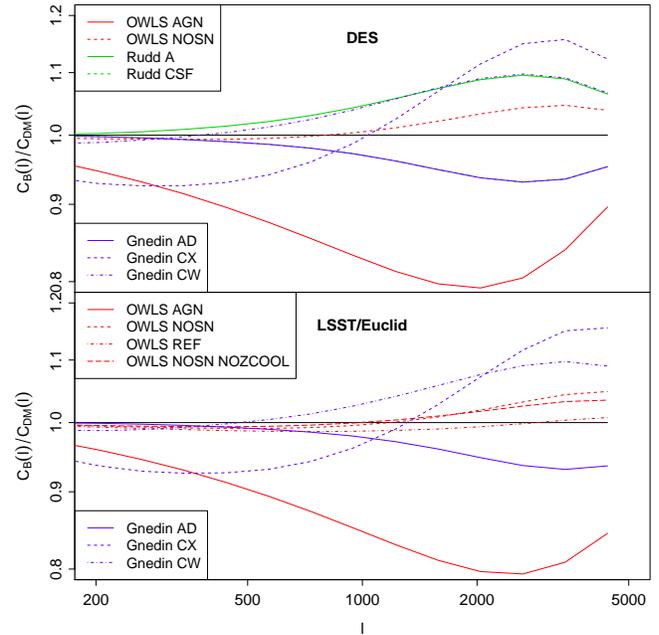}
    \caption{The ratio of shear tomography power spectra of different baryonic scenarios with respect to the dark matter only scenario for the lowest auto-correlation tomography bin.} \label{fi:cl_diff}
\end{figure}
\subsection{Projected shear power spectra from the baryonic scenarios}
\label{sec:data}

The three sets of simulations described in Sect. \ref{sec:sim} have different input cosmologies. In order to create a coherent set of baryonic scenarios we assume that the cosmology dependence enters through the dark matter power spectrum only and ``re-normalize'' the 3D density power spectra for each baryonic scenario via    
\be 
\label{eq:pdelta1}
P^\mr{bary, theory}_\delta (k,z) =  \frac{P^\mr{bary, sim}_\delta (k,z)}{P^\mr{DM, sim}_\delta (k,z)} P^\mr{DM, theory}_\delta (k,z)
\ee
where $P^\mr{bary, sim}_\delta (k,z)$ denotes the joint dark+baryonic power spectrum from a given simulation, $P^\mr{DM, sim}_\delta$ is the corresponding dark matter only power spectrum, and $P^\mr{DM, theory}_\delta$ is the dark matter power spectrum calculated from \textsc{CosmoLike} (see Sect. \ref{sec:model} for details) assuming a Planck+WMAP polarization best-fit cosmology.

In each case, the simulations treat finite volumes and have limited resolutions, so the simulated spectra alone do not suffice to cover the entire range of wave numbers needed. As such, it is necessary to extrapolate simulation results using a particular theoretical model. The Rudd et al. (2008) simulations pose the most stringent constraints on the range of $k$ and $z$, i.e. matter power spectra are accurate over a range of $k \in[0.3;10]\, h\mr{Mpc^{-1}}$, where the lower $k$-limit is a consequence of simulation size, and over a range of $z \in [0.0;2.0]$. Outside the $k$-ranges we extrapolate $P_\delta$ with a theoretical power spectrum (see below); however, note that the limited redshift range of the Rudd et al. simulations prohibits us from computing LSST shear power spectra because the LSST redshift range extends to $z=3.5$. Overall this give us 14 baryonic scenarios for the DES survey and 12 for LSST.

Having obtained the density power spectra we calculate the shear power spectra as
\begin{equation}
\label{eq:pdeltatopkappa}
C ^{ij} (l) = \frac{9H_0^4 \om^2}{4c^4} \int_0^{\chi_\mr h} 
\mr d \chi \, \frac{g^{i}(\chi) g^{j}(\chi)}{a^2(\chi)} \pd \left(\frac{l}{f_K(\chi)},\chi \right) \,,
\end{equation}
with $l$ being the 2D wave vector perpendicular to the line of sight,
$\chi$ denoting the comoving coordinate, $\chi_\mr h$ is the comoving
coordinate of the horizon, $a(\chi)$ is the scale factor, and
$f_K(\chi)$ the comoving angular diameter distance (throughout set to $\chi$ since we assume a flat Universe).
The lens efficiency $g^{i}$ is defined as an integral over the
redshift distribution of source galaxies $n(\chi(z))$ in the
$i^\mr{th}$ tomographic interval
\begin{equation}
\label{eq:redshift_distri}
g^{i}(\chi) = \int_\chi^{\chi_{\mr h}} \mr d \chi' n^{i} (\chi') \frac{f_K (\chi'-\chi)}{f_K (\chi')} \,.
\end{equation}

In this analysis we use two different redshift distributions mimicking a DES and and LSST/Euclid like survey and divide each redshift range into five bins (see Fig. \ref{fi:zdistrib} and Table \ref{tab:survey}). For LSST we adopt the redshift distribution suggested in \cite{cjj13} and the DES redshift distribution is modeled by a modified CFHTLS redshift distribution \citep[see][adjusted for the slightly lower mean redshift of DES]{bhs07}. The exact parameterization for the latter reads
\be 
\label{redshiftben}
n(z)=N \, \left( \frac{z}{z_0}\right)^\alpha \exp \left[ - \left(  \frac{z}{z_0} \right)^\beta \right]\,,
\ee 
with $\alpha=2.0$, $\beta=1.0$, $z_0=0.5$.

Since we chose five tomographic bins, the resulting data vector which enters the likelihood analysis consists of 15 tomographic shear power spectra, each with 20 logarithmically spaced bins ($l \in [30;5000]$), hence 300 data points overall. The limits of the tomographic $z$-bins are chosen such that each bin contains a similar number of galaxies.

In Fig. \ref{fi:cl_final} we show the uncertainty range spanned by the baryonic scenarios (grey shaded area) with respect to the DM only scenario (black line) for the 5 auto-correlation redshift shear power spectra. In Fig. \ref{fi:cl_diff} we further show the ratio of baryonic to dark matter $C^{11}(l)$ shear power spectrum for a subset of the scenarios. One can clearly see that at different $l$ the range is bracketed by different scenarios, with the strong AGN-feedback scenario being the lower extreme starting from $l\sim400$ and the extreme cooling scenario (CX) being upper limit for $l>2000$.

\section{Likelihood analysis: neglecting baryons}
\label{sec:like_without}
We first carry out likelihood analyses with shear tomography power spectra from the various baryonic scenarios as the input data vectors without accounting for baryons, i.e. using the DM power spectrum in the model vector only. 

\begin{table*}
\caption{Fiducial cosmology, minimum and maximum of the flat prior on cosmological parameters, and Planck prior information used in the analysis. }
\begin{center}
\begin{tabular}{|l|c c c c c c c|}
\hline
\hline
&$\om$ & $\sig$ & $\ns$ & $\w$ & $\wa$ & $\omb$ & $h_0$ \\
\hline
Fiducial & 0.315 & 0.829 & 0.9603 & -1.0 & 0.0 & 0.049 & 0.673\\
Min & 0.1 & 0.6 & 0.85 & -2.0 & -2.5 & 0.04 & 0.6\\
Max & 0.6 & 0.95 & 1.06 & 0.0 & 2.5 & 0.055 & 0.76\\
Planck+WP 1-$\sigma$ & $^{+0.016}_{-0.018}$ & $\pm 0.012$ & $\pm 0.0073$ & - & - & $\pm 0.00062$ & $\pm 0.012$\\
\hline
\hline
\end{tabular}
\end{center}
\label{tab:cosmology}
\end{table*}

\begin{figure*}
\includegraphics[width=17cm]{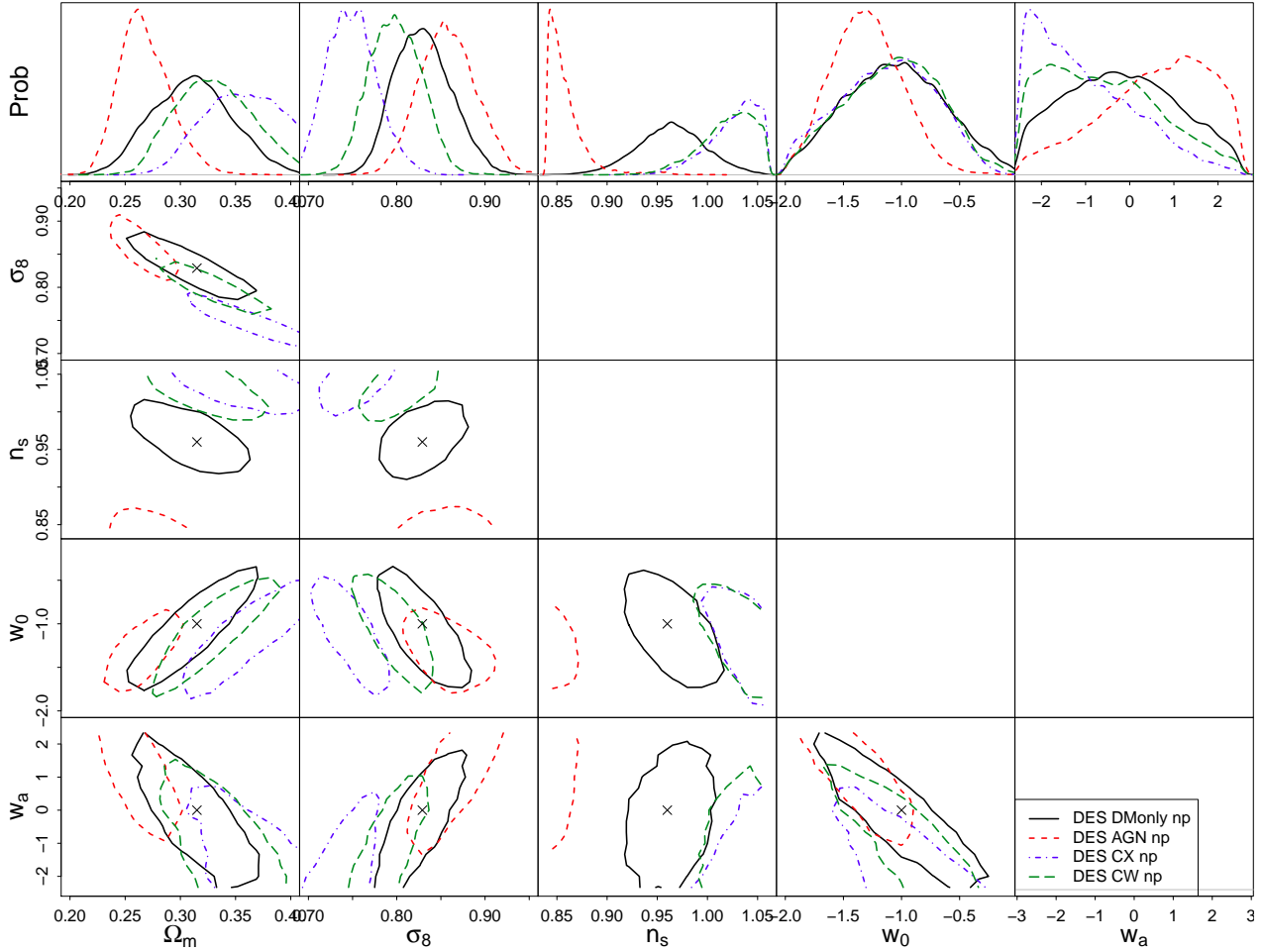}
\caption{Cosmological constraints for a DES survey assuming different underlying baryonic scenarios for our Universe, i.e. pure dark matter (\ti{black/solid}), strong AGN feedback (\ti{red/dashed}), extreme cooling (\ti{blue/dot-dashed}), and moderate cooling (\ti{green/long-dashed}), which are unaccounted for in the likelihood analysis. The scenarios are detailed in Sect. \ref{sec:sim}. The characters ``np" labeling each model indicate that the analysis is performed with no priors on the parameters.}
         \label{fi:des_np}
\end{figure*}

\begin{figure*}
\includegraphics[width=17cm]{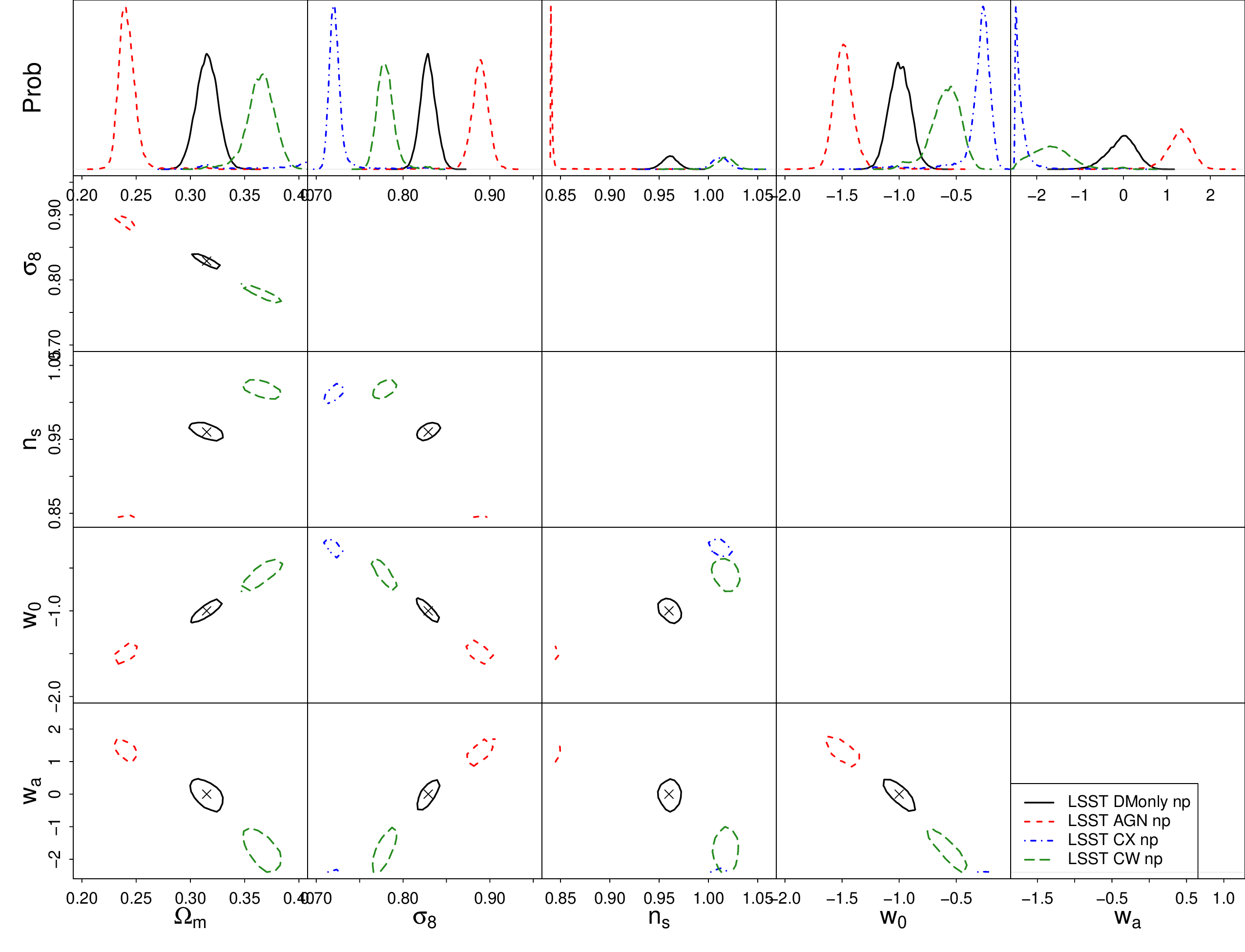}
\caption{Cosmological constraints for a LSST/Euclid survey assuming different underlying baryonic scenarios for our Universe, i.e. pure dark matter (\ti{black/solid}), strong AGN feedback (\ti{red/dashed}), extreme cooling (\ti{blue/dot-dashed}), and moderate cooling (\ti{green/long-dashed}), which are unaccounted for in the likelihood analysis. The scenarios are detailed in Sect. \ref{sec:sim}.}
         \label{fi:lsst_np}
\end{figure*}

\subsection{Modeling Cosmological Quantities}
\label{sec:model}
\paragraph*{Shear tomography power spectra} All simulated likelihood analyses in this paper are computed using the weak lensing modules of  \textsc{CosmoLike} \citep[see][for an early version; official release paper is Krause et al. 2014 in prep]{eks14}. We compute the linear power spectrum using the \cite{eh99} transfer function and model the non-linear evolution of the density field as described in \cite{tsn12}.
Time-dependent dark energy models ($w=w_0+(1-a)\,w_a$) are incorporated following the recipe of {\sc icosmo} \citep{rak11}, which in the non-linear regime interpolates Halofit between flat and open cosmological models \citep[also see][for more details]{shj10}. From the density power spectrum we compute the shear power spectrum as described in Sect. \ref{sec:data}.

\paragraph*{Shear covariances} Under the assumption that the 4pt-function of the shear field can be expressed in terms of 2pt-functions (so-called Gaussian shear field) the covariance of projected shear power spectra can be calculated as in \citep{huj04} 
\begin{widetext}
\be
\label{eq:covhujain}
\mr{Cov_G} \left( C^{ij} (l_1) C^{kl} (l_2) \right) = \langle \Delta C^{ij} (l_1) \, \Delta C^{kl} (l_2) \rangle  =  \frac{\delta_{l_1 l_2}}{ 2 f_\mr{sky} l_1 \Delta l_1}  \left[\bar C^{ik}(l_1) \bar C^{jl}(l_1) + \bar C^{il}(l_1) \bar C^{jk} (l_1) \right]\,,
\ee
\end{widetext}
with
\be
\label{details}
\bar C^{ij}(l_1)= C^{ij}(l_1)+ \delta_{ij} \frac{\sigma_\eps^2}{n^{i}} \,,
\ee
where the superscripts indicate the redshift bin; $n^{i}$ is the density of source galaxies in the $i$-th redshift bin; and $\sigma_\eps$ is the RMS of the shape noise.

Since non-linear structure growth at late time induces significant non-Gaussianities in the shear field, using the covariance of Eq.~(\ref{eq:covhujain}) in a likelihood analysis results in underestimates of the errors on cosmological parameters. Therefore, the covariance must be amended by an additional term, i.e. $\mr{Cov}=\mr{Cov_G}+\mr{Cov_{NG}}$.  The non-Gaussian covariance is calculated from the convergence trispectrum $T_{\kappa}$ \citep{CH01,taj09}, and we include a sample variance term $T_{\kappa,\rm{HSV}}$ that describes scatter in power spectrum measurements due to large scale density modes \citep{tb07, sht09},
\begin{widetext}
\be
 \mr{Cov_{NG}}(C^{ij}(l_1),C^{kl}(l_2)) =  \int_{|\mathbf l|\in l_1}\frac{d^2\mathbf l}{A(l_1)}\int_{|\mathbf l'|\in l_2}\frac{d^2\mathbf l'}{A(l_2)} \left[\frac{1}{\Omega_{\mr s}}T_{\kappa,0}^{ijkl}(\mathbf l,-\mathbf l,\mathbf l',-\mathbf l') + T_{\kappa,\rm{HSV}}^{ijkl}(\mathbf l,-\mathbf l,\mathbf l',-\mathbf l') \right] \,,
\ee
\end{widetext}
with $A(l_i) = \int_{|\mathbf l|\in l_i}d^2\mathbf l \approx 2 \pi l_i\Delta l_i$ the integration area associated with a power spectrum bin centered at $l_i$ and width $\Delta l_i$.

The convergence trispectrum $T_{\kappa,0}^{ijkl}$ is, in the absence of finite volume effects, defined as  
\begin{widetext}
\be
\label{eq:tri2}
T_{\kappa,0}^{ijkl} (\mathbf l_1,\mathbf l_2,\mathbf l_3,\mathbf l_4) = \left( \frac{3}{2} \frac{H_0^2}{c^2} \om \right)^{4} \int_0^{\chi_h} \d \chi \, \left( \frac{\chi}{a(\chi)}\right)^4  g^i g^j g^k g^l \times \chi^{-6} \, T_{\delta,0}  \left( \frac{\mathbf l_1}{\chi}, \frac{\mathbf l_2}{\chi}, \frac{\mathbf l_3}{\chi}, \frac{\mathbf l_4}{\chi}, z(\chi) \right) \,,
\ee
\end{widetext}
with $T_{\delta,0}$ the matter trispectrum (again, not including finite volume effects), and where we abbreviated $g^i=g^i(\chi)$.

We model the matter trispectrum using the halo model \citep{Seljak00, CS02}, which assumes that all matter is bound in virialized structures that are modeled as biased tracers of the density field. Within this model the statistics of the density field can be described by the dark matter distribution within halos on small scales, and is dominated by the clustering properties of halos and their abundance on large scales. In this model, the trispectrum splits into five terms describing the 4-point correlation within one halo (the \emph{one-halo} term $T^{\mr{1h}}$), between 2 to 4 halos (\emph{two-, three-, four-halo} term), and a so-called halo sample variance term $T_{\mr{HSV}}$, caused by fluctuations in the number of massive halos within the survey area,
\be
\label{eq:t}
T = T_0 + T_{\mr{HSV}} = \left[T_{\mr{1h}}+T_{\mr{2h}}+T_{\mr{3h}}+T_{\mr{4h}}\right]+T_{\mr{HSV}}\;.
\ee
The \emph{two-halo} term is split into two parts, representing correlations between two or three points in the first halo and two or one point in the second halo. As halos are the building blocks of the density field in the halo approach, we need to choose models for their internal structure, abundance and clustering in order to build a model for the trispectrum.

Our implementation of the one-, two- and four-halo term contributions to the matter trispectrum follows \citet{CH01}, and we neglect the three-halo term as it is subdominant compared to the other terms at the scales of interest for this analysis. Specifically, we assume NFW halo profiles \citep{NFW} with the \citet{Bhattacharya11} fitting formula for the halo mass--concentration relation $c(M,z)$, and the \citet{Tinker10} fit functions for the halo mass function $\frac{ dn}{dM}$ and linear halo bias $b(M)$ (all evaluated at $\Delta = 200$), neglecting terms involving higher order halo biasing.

Within the halo model framework, the halo sample variance term is described by the change of the number of massive halos within the survey area due to survey-scale density modes; following \citet{sht09} it is calculated as
\begin{widetext}
\bea
T_{\kappa,\rm{HSV}}^{ijkl}(\mathbf l_1,-\mathbf l_1,\mathbf l_2,-\mathbf l_2)= \left(\frac{3}{2}\frac{H_0^2}{c^2}\Omega_{\mr m}\right)^4 &\times&  \int_0^{\chi_\mr h}d\chi \left(\frac{d^2 V}{d\chi d\Omega}\right)^2 \left(\frac{\chi}{a(\chi)}\right)^4 g^i g^j g^k g^l \nn \\
&\times&  \int d M \frac{d n}{d M} b(M)\left(\frac{M}{\bar{\rho}}\right)^2 |\tilde{u}(l_1/\chi, c(M,z(\chi))|^2 \nn \\
 &\times& \int d M' \frac{d n}{d M'} b(M')\left(\frac{M'}{\bar{\rho}}\right)^2 |\tilde{u}(l_2/\chi, c(M',z(\chi))|^2 \nn \\
 &\times&  \int_0^\infty \frac{k dk}{2\pi}P_\delta^{\mr{lin}}(k,z(\chi))|\tilde W(k\chi \Theta_{\mr s})|^2 \,.
\eea
\end{widetext}

\subsection{Likelihood Analysis without PCA mitigation of baryons}
\label{sec:results1}

We have introduced the mathematical basics of likelihood analyses in Sect. \ref{sec:likebasics} and the \textsc{CosmoLike} internal calculation of our data vectors, model vectors, and covariances in Sect. \ref{sec:data} and Sect. \ref{sec:model}. \textsc{CosmoLike} samples the parameter space using a parallel MCMC of \cite{emcee} algorithm implemented through the python emcee package\footnote{http://dan.iel.fm/emcee/current/user/pt/} \citep{fhg13}. Altogether we present results of 52 simulated likelihood analyses in this paper; each analysis consists of 108,000 MCMC steps (after discarding 12000 steps as burn-in phase) in a seven dimensional cosmological parameter space with flat priors at the boundaries of the parameter range (see Table \ref{tab:cosmology}). We check for convergence by running several shorter chains for all scenarios and ten chains with 480000 MCMC steps and find no qualitative change in the contours. 

We have analyzed all baryonic scenarios described in Sect. \ref{sec:sim}, but confine our detailed results to two extreme scenarios (AGN, CX) and two moderate scenarios (AD, CW). We run analyses for a DES and LSST/Euclid survey without prior information (except for the flat priors at the limits of our parameter space); results for the same analysis with prior information from the Planck mission can be found in Appendix \ref{sec:app}. All contour plots are marginalized over five cosmological parameters; in addition to the ones mentioned in the plots we marginalize over $\omb$ and $H_0$. The first row of all figures with contour plots show the posterior probability distribution of a given cosmological parameter marginalized over the other six cosmological parameters.

Figures~\ref{fi:des_np}  and \ref{fi:lsst_np} compare the impact of strong AGN feedback (AGN, \ti{dashed red}), extreme cooling (CX,  \ti{dashed-dotted blue}), moderate cooling (CW, \ti{long-dashed green} ), to the DM scenario (\ti{black solid}) for DES and LSST/Euclid, respectively\footnote{All contours shown in this paper indicate the $68\%$ confidence regions.}. When baryons are not accounted for, the parameter estimates are severely biased. We quantify these biases by showing the marginalized 1D best-fit cosmological parameters and their 1-$\sigma$ error bars in Tables \ref{tab:paradist_DESnp} and \ref{tab:paradist_LSSTnp} (see rows with PCA order$=0$). 

Note the extremely large biases in Fig.~\ref{fi:lsst_np}. For example, the best fit value of $w_0$ if the baryons behaved as in the CX scenario would be $-0.316$, differing from the ``true'' value of $w=-1$ by 0.684, or almost 6-$\sigma$. This effect is even more significant for the AGN scenario. As a side-note we point out that quoting a bias as multiples of $\sigma$ assumes the posterior probability to be Gaussian, which is done implicitly in all Fisher analyses of previous papers. Looking at the 1D posterior probabilities in Figs. \ref{fi:des_np}, \ref{fi:lsst_np}, \ref{fi:des_pp}, \ref{fi:lsst_pp}, this is hardly justified; all posteriors show a substantial skewness or kurtosis. As a consequence a quantitative comparison to previous, similar analyses that are based on Fisher matrices is not meaningful.
 
In any case, Fisher matrix or MCMC, it has become clear that neglecting the effects of baryons would lead to a catastrophic mis-interpretation of the data and a mitigation strategy is essential for Stage IV surveys. Given the significantly larger statistical error bars expected in DES, the resulting bias in Fig. \ref{fi:des_np} is less severe than for the LSST case, nevertheless, even for DES a mitigation scheme for baryons is necessary.

\section{PCA Marginalization over baryonic uncertainties}
\label{sec:like}

\subsection{Identifying the Principal Components}
\label{sec:identPC}
\begin{figure*}
\includegraphics[width=18cm]{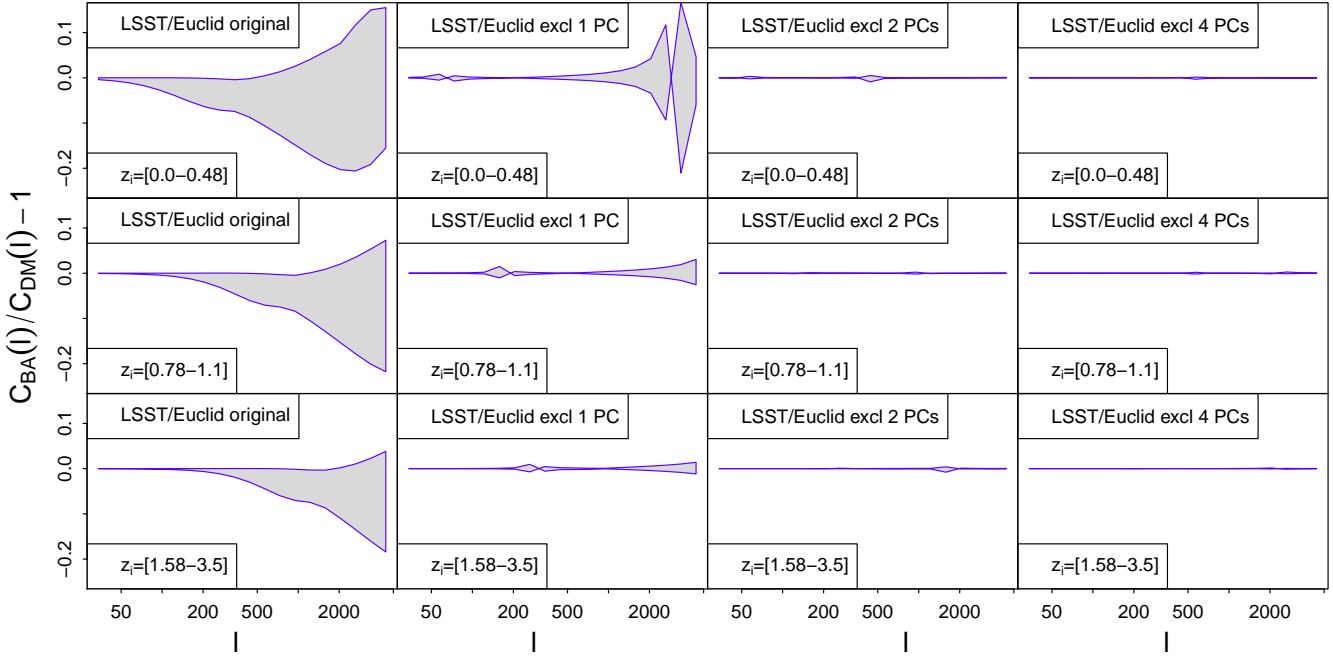}
\caption{This plot shows the uncertainty range spanned by the baryonic scenarios, centralized around the DM scenario, when excluding one (\ti{second panel}) two (\ti{third panel}), and four (\ti{fourth panel}) PCs compared to the original uncertainty range (\ti{left panel}). The three panel rows show three tomographic autocorrelation power spectra for LSST/Euclid.}
         \label{fi:powspec}
\end{figure*}

\begin{table*}
\caption{Projection angle of the difference vectors ($C_\mr{bary}^{ij}(l) - C_\mr{DM}^{ij}(l)$) onto the PCs (see Eq. \ref{eq:angle}) and fraction of this difference vector that is captured by the PC subspace (see Eq. \ref{eq:vol}).}
\begin{center}
\begin{tabular}{|l|r c|r c|r c|r c|r c|r c|r c|r c|}
\hline
&\multicolumn{8}{|c|}{DES}&\multicolumn{8}{|c|}{LSST} \\ \hline
Baryonic Scenario & $\abs{\cos \theta_1}$ & $V_1$ & $\abs{\cos \theta_2}$ & $V_2$& $\abs{\cos \theta_3}$ & $V_3$ &$\abs{\cos \theta_4}$ & $V_4$ &$\abs{\cos \theta_1}$ &$V_1$ &$\abs{\cos \theta_2}$ &$V_2$ &$\abs{\cos \theta_3}$ & $V_3$&$\abs{\cos \theta_4}$ & $V_4$ \\ \hline
$\mr{AGN}$ & 0.98 & 0.98 & 0.17 & 1 & 0.002 & 1 & 0.0097 & 1 & 0.95 & 0.95 & 0.31 & 1 & 0.026 & 1 & 0.00056 & 1 \\
$\mr{NOSN}$ & 0.87 & 0.87 & 0.47 & 0.99 & 0.11 & 1 & 0.047 & 1 & 0.97 & 0.97 & 0.1 & 0.98 & 0.052 & 0.98 & 0.21 & 1 \\
$\mr{NOSN\,NOZCOOL}$ & 0.88 & 0.88 & 0.46 & 1 & 0.087 & 1 & 0.04 & 1 & 0.96 & 0.96 & 0.18 & 0.98 & 0.06 & 0.98 & 0.18 & 1 \\
$\mr{NOZCOOL}$ & 0.43 & 0.43 & 0.86 & 0.96 & 0.001 & 0.96 & 0.27 & 1 & 0.99 & 0.99 & 0.085 & 0.99 & 0.078 & 1 & 0.051 & 1 \\
$\mr{REF}$ & 0.63 & 0.63 & 0.77 & 1 & 0.09 & 1 & 0.03 & 1 & 0.99 & 0.99 & 0.097 & 1 & 0.05 & 1 & 0.048 & 1 \\
$\mr{WDENS}$ & 0.99 & 0.99 & 0.12 & 1 & 0.018 & 1 & 0.024 & 1 & 0.88 & 0.88 & 0.44 & 0.99 & 0.14 & 1 & 0.0074 & 1 \\
$\mr{DBLIMFV1618}$ & 0.99 & 0.99 & 0.13 & 1 & 0.003 & 1 & 0.0065 & 1 & 0.95 & 0.95 & 0.31 & 1 & 0.031 & 1 & 0.0058 & 1 \\
$\mr{WML4}$ & 0.61 & 0.61 & 0.78 & 0.99 & 0.05 & 0.99 & 0.14 & 1 & 0.99 & 0.99 & 0.069 & 1 & 0.06 & 1 & 0.037 & 1 \\
$\mr{WML1V848}$ & 0.98 & 0.98 & 0.21 & 1 & 0.012 & 1 & 0.025 & 1 & 0.97 & 0.97 & 0.26 & 1 & 0.025 & 1 & 0.005 & 1 \\
$\mr{AD}$ & 0.98 & 0.98 & 0.21 & 1 & 0.056 & 1 & 0.013 & 1 & 0.3 & 0.3 & 0.95 & 0.99 & 0.002 & 0.99 & 0.086 & 1 \\
$\mr{CX}$ & 0.76 & 0.76 & 0.64 & 1 & 0.015 & 1 & 0.00035 & 1 & 0.99 & 0.99 & 0.16 & 1 & 0.0015 & 1 & 0.00077 & 1 \\
$\mr{CW}$ & 0.97 & 0.97 & 0.23 & 1 & 0.014 & 1 & 0.0078 & 1 & 0.87 & 0.87 & 0.49 & 1 & 0.03 & 1 & 0.031 & 1 \\
$\mr{A}$ & 1 & 1 & 0.079 & 1 & 0.036 & 1 & 0.026 & 1 & -- & -- & -- & -- & -- & -- & -- & -- \\
$\mr{CSF}$ & 0.98 & 0.98 & 0.2 & 1 & 0.032 & 1 & 0.006 & 1 & -- & -- & -- & -- & -- & -- & -- & -- \\
\hline
\end{tabular}
\end{center}
\label{tab:barypcasub}
\end{table*}

Recall that the PCA marginalization scheme as outlined in Sect. \ref{sec:likebasics} starts with creating a set of model vectors at each point in cosmology that spans the variation under nuisance parameters. This ideal case corresponds to having a representative set of simulated baryonic scenarios at each point in cosmology, which unfortunately is computationally unfeasible. Here we rely on the approximation we already detailed in Sect. \ref{sec:data}, namely that the cosmology enters through the dark matter power spectrum only. 

Following Eqs. (\ref{eq:pdelta1}, \ref{eq:pdeltatopkappa}), we compute the baryonic shear power spectrum at any given cosmology $\pco$ from the set of baryonic shear power spectra we computed in Sect. \ref{sec:data} for the fiducial cosmology $\pco^\mr{fid}$ as
\be 
\label{eq:pdelta2}
C^{ij}_\mr{bary} (l,\pco) =  \frac{C^{ij}_\mr{bary} (l, \pco^\mr{fid})}{C^{ij}_\mr{DM} (l, \pco^\mr{fid})} C^{ij}_\mr{DM} (l,\pco)
\ee
where $C^{ij}_\mr{DM} (l,\pco)$ is computed from \textsc{CosmoLike}.

For each point in parameter space sampled in the MCMC, we use Eq.~(\ref{eq:pdelta2}) to compute 14 (12) baryonic shear power spectra for DES (LSST/Euclid). We concatenate the shear power spectra to a $300\times14$ ($300\times12$ for LSST/Euclid) matrix, which defines the set of model vectors $\vek M_\alpha$ that is assumed to span the uncertainty due to baryons. We can now define the difference matrix $\Del$ as in  Eq. (\ref{eq:diffmat}) and perform a (full) SVD on this matrix using Eq. (\ref{eq:SVD}), which gives the transformation matrix $\U$, with the principal components as columns. One at a time, we remove the PCs with the largest singular values.

This gives us the necessary ingredients to continue with the procedure outlined in Sect. \ref{sec:likebasics}. Figure \ref{fi:powspec} shows the envelope of the different baryonic simulations for three different auto-redshift power spectra (corresponding to the three rows), and each column depicts the result of removing more modes. The first column shows the uncertainties from baryonic physics if no modes were removed. The second column shows that even by removing only a single mode, we are able to reduce the baryonic uncertainties by a significant amount. Removing 4 modes seems to remove any lingering ambiguity associated with the baryons. This is a striking result: by throwing away only less than 2\% of the data (4 modes out of 300), we have created a ``baryon-free'' subset that can be analyzed with the dark matter power spectrum.

In addition to the analysis in Fig. \ref{fi:powspec} we determine the number of PCs by examining the projections of difference vectors $\vek\Delta_\alpha$ onto the PC subspaces. Recall that for each baryonic scenario $\alpha$ we calculate a difference vector $\vek\Delta_\alpha$. We can project each of these vectors onto the subspace spanned by the PC modes that are removed. In particular we compute the absolute value of the cosine of the projection angle 
\be
\label{eq:angle}
\cos \theta_i^\alpha =  \frac{\vek \Delta_\alpha \cdot \vek{PC_i}}{|\vek \Delta_\alpha||\vek{PC_i}|} \,, 
\ee
between the $\alpha$-th difference vector and $i$-th PC. The corresponding PC captures all baryonic uncertainty of scenario $\alpha$ if $\abs{\cos \theta_i^\alpha}=1$ and none if $\abs{\cos \theta_i^\alpha}=0$.
When removing $n$ PCs we can define the fraction of the difference vector that falls into the space spanned by the PCs as 
\be
\label{eq:vol}
V_n = \sqrt{\sum_i^n \cos^2 \theta_i^\alpha} \, .
\ee
Table \ref{tab:barypcasub} show $\theta_i^\alpha$ and $V_n$ for all the simulations. Even removing two modes seems almost sufficient to remove the differences caused by baryonic effects. This analysis shows impressively that the baryonic scenarios for both DES and LSST/Euclid are almost completely captured within PCA subspaces of relatively low dimensionality; when using a four-dimensional PC space the worst scenarios is still to $99.5\%$ per cent within the PCA-volume.

\subsection{Results of the likelihood analyses}
\label{sec:results2}

\begin{figure*}
\includegraphics[width=18cm]{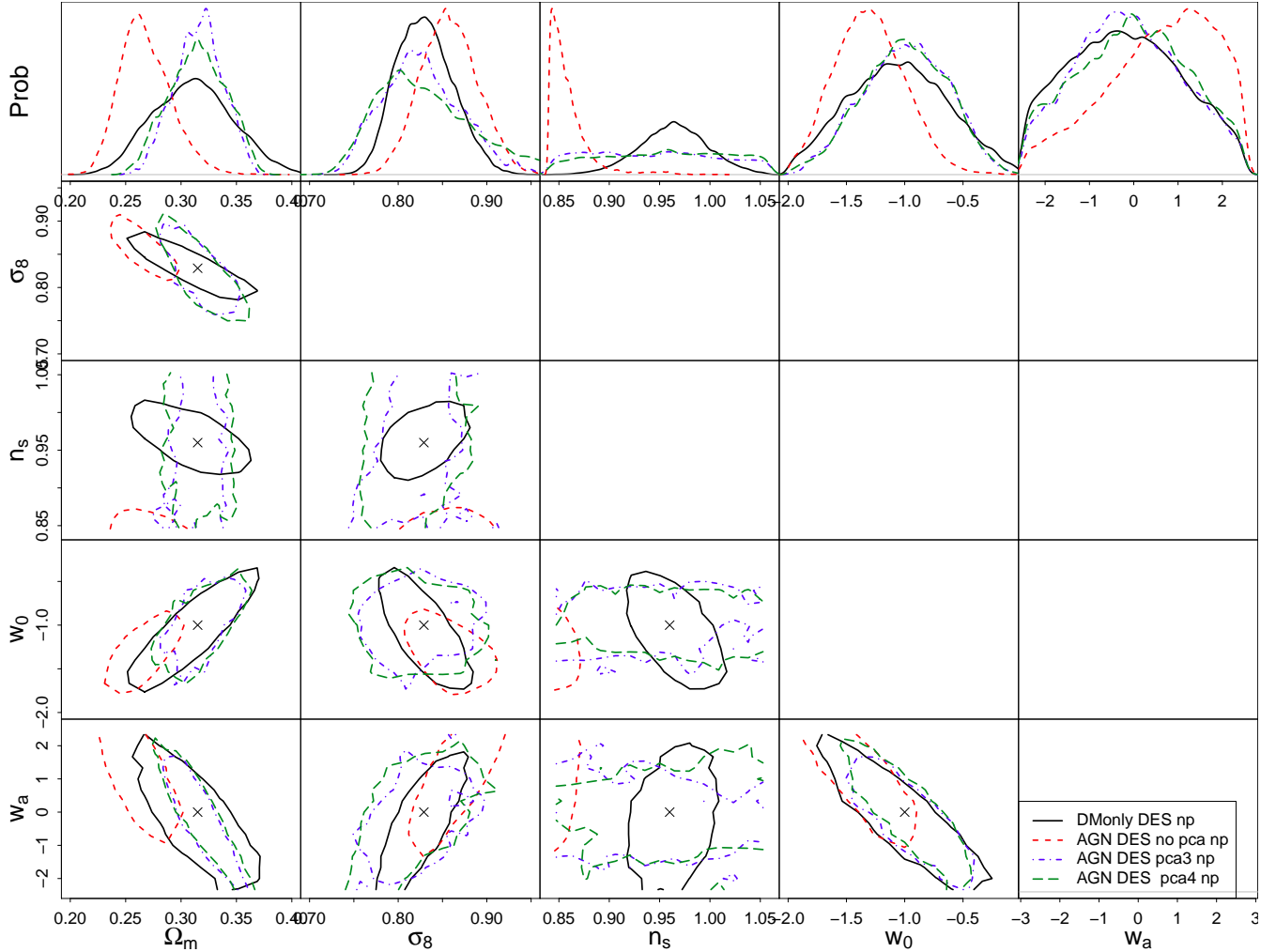}
\caption{Cosmological constraints for a DES survey when using the PCA mitigation technique. The results shown assume that the baryonic physics of the Universe follows the AGN scenario (i.e. the most extreme baryonic scenario). We remove three and four PC modes (\ti{blue/dashed} and \ti{green/long-dashed}, respectively) and compare the results to the untreated AGN scenario (\ti{red/dashed}) and to a pure DM scenario (\ti{black/solid}).}
         \label{fi:des_pca_np}
\end{figure*}

\begin{table*}
\caption{Marginalized 1D constraints on cosmological parameters for the DM, AD, AGN, CW, and CX scenario with and without the PCA mitigation for a DES survey (no priors). The last column contains the $\Delta \chi^2$ distance (see Eq. \ref{eq:paradistance}) between best fit and fiducial parameter point.}
\begin{center}
\def\arraystretch{1.3}
\begin{tabular}{|c c c c c c c c c c|}
\hline
Scenario & PCA order & $\om$ & $\sig$ & $\ns$ & $\w$ &$\wa$ & $\omb$ & $h_0$ & $\Delta \chi^2$ \\
\hline
DM&0&0.311$_{-0.0373}^{+0.0363}$ &0.83$_{-0.0324}^{+0.0328}$ &0.964$_{-0.0334}^{+0.0333}$ &-1.06$_{-0.448}^{+0.438}$
&-0.184$_{-1.37}^{+1.36}$
&0.0474$_{-0.00517}^{+0.00516}$
&0.691$_{-0.0763}^{+0.0786}$
&0.675  \\ 
&&&&&&&&&\\ 
AD&0&0.299$_{-0.0346}^{+0.0351}$ &0.849$_{-0.0324}^{+0.0332}$ &0.932$_{-0.0353}^{+0.0348}$ &-1.05$_{-0.426}^{+0.438}$
&0.106$_{-1.37}^{+1.38}$
&0.0475$_{-0.00502}^{+0.005}$
&0.735$_{-0.0795}^{+0.0827}$
&3.57  \\ 
AD&3&0.317$_{-0.0234}^{+0.0238}$ &0.818$_{-0.04}^{+0.0416}$ &0.941$_{-0.0724}^{+0.0752}$ &-1.03$_{-0.355}^{+0.352}$
&-0.234$_{-1.41}^{+1.39}$
&0.0476$_{-0.00495}^{+0.00506}$
&0.682$_{-0.0717}^{+0.0722}$
&0.574  \\ 
AD&4&0.311$_{-0.0245}^{+0.0257}$ &0.832$_{-0.0556}^{+0.0568}$ &0.951$_{-0.0696}^{+0.0705}$ &-1.07$_{-0.366}^{+0.356}$
&0.0698$_{-1.4}^{+1.38}$
&0.0473$_{-0.00484}^{+0.005}$
&0.685$_{-0.076}^{+0.0795}$
&0.784  \\ 
&&&&&&&&&\\ 
AGN&0&0.268$_{-0.0224}^{+0.0228}$ &0.858$_{-0.0324}^{+0.0326}$ &0.86$_{-0.0157}^{+0.0131}$ &-1.3$_{-0.306}^{+0.302}$
&0.579$_{-1.31}^{+1.27}$
&0.0463$_{-0.00468}^{+0.00497}$
&0.797$_{-0.0753}^{+0.074}$
&55.5  \\ 
AGN&3&0.316$_{-0.0225}^{+0.0225}$ &0.827$_{-0.0453}^{+0.0467}$ &0.944$_{-0.0703}^{+0.0717}$ &-0.996$_{-0.386}^{+0.378}$
&-0.12$_{-1.24}^{+1.28}$
&0.0474$_{-0.00499}^{+0.00495}$
&0.686$_{-0.0764}^{+0.0729}$
&0.702  \\ 
AGN&4&0.315$_{-0.027}^{+0.0272}$ &0.833$_{-0.0557}^{+0.0584}$ &0.955$_{-0.0713}^{+0.0696}$ &-1$_{-0.374}^{+0.378}$
&-0.0381$_{-1.37}^{+1.33}$
&0.0481$_{-0.00533}^{+0.00494}$
&0.684$_{-0.0797}^{+0.0803}$
&0.599  \\ 
&&&&&&&&&\\ 
CW&0&0.332$_{-0.0366}^{+0.0363}$ &0.799$_{-0.0297}^{+0.0298}$ &1.02$_{-0.0272}^{+0.0273}$ &-1.07$_{-0.438}^{+0.413}$
&-0.579$_{-1.34}^{+1.33}$
&0.0477$_{-0.00508}^{+0.00501}$
&0.61$_{-0.067}^{+0.0668}$
&13  \\ 
CW&3&0.317$_{-0.0212}^{+0.0224}$ &0.822$_{-0.0353}^{+0.0379}$ &0.951$_{-0.0746}^{+0.0731}$ &-1.03$_{-0.351}^{+0.353}$
&-0.0789$_{-1.33}^{+1.31}$
&0.0478$_{-0.00516}^{+0.00497}$
&0.675$_{-0.0712}^{+0.0697}$
&0.421  \\ 
CW&4&0.316$_{-0.0234}^{+0.0243}$ &0.826$_{-0.0512}^{+0.0535}$ &0.956$_{-0.0739}^{+0.0699}$ &-1.02$_{-0.363}^{+0.372}$
&-0.107$_{-1.4}^{+1.36}$
&0.0472$_{-0.00495}^{+0.00501}$
&0.674$_{-0.0786}^{+0.0808}$
&0.601  \\ 
&&&&&&&&&\\ 
CX&0&0.364$_{-0.0415}^{+0.0413}$ &0.749$_{-0.0284}^{+0.028}$ &1.03$_{-0.0255}^{+0.0247}$ &-1.13$_{-0.44}^{+0.425}$
&-1$_{-1.16}^{+1.26}$
&0.0477$_{-0.00514}^{+0.00509}$
&0.551$_{-0.0631}^{+0.0629}$
&32.7  \\ 
CX&3&0.315$_{-0.0228}^{+0.0229}$ &0.816$_{-0.0389}^{+0.0401}$ &0.947$_{-0.0726}^{+0.0713}$ &-1.1$_{-0.35}^{+0.348}$
&-0.107$_{-1.39}^{+1.33}$
&0.0474$_{-0.00517}^{+0.00518}$
&0.681$_{-0.0702}^{+0.0735}$
&0.822  \\ 
CX&4&0.314$_{-0.0257}^{+0.0253}$ &0.822$_{-0.0529}^{+0.0537}$ &0.945$_{-0.0704}^{+0.0719}$ &-1.07$_{-0.34}^{+0.354}$
&-0.0428$_{-1.45}^{+1.41}$
&0.0477$_{-0.00509}^{+0.005}$
&0.684$_{-0.0823}^{+0.0844}$
&1.01  \\ 
\hline
\end{tabular}
\end{center}
\label{tab:paradist_DESnp}
\end{table*}

\begin{figure*}
\includegraphics[width=18cm]{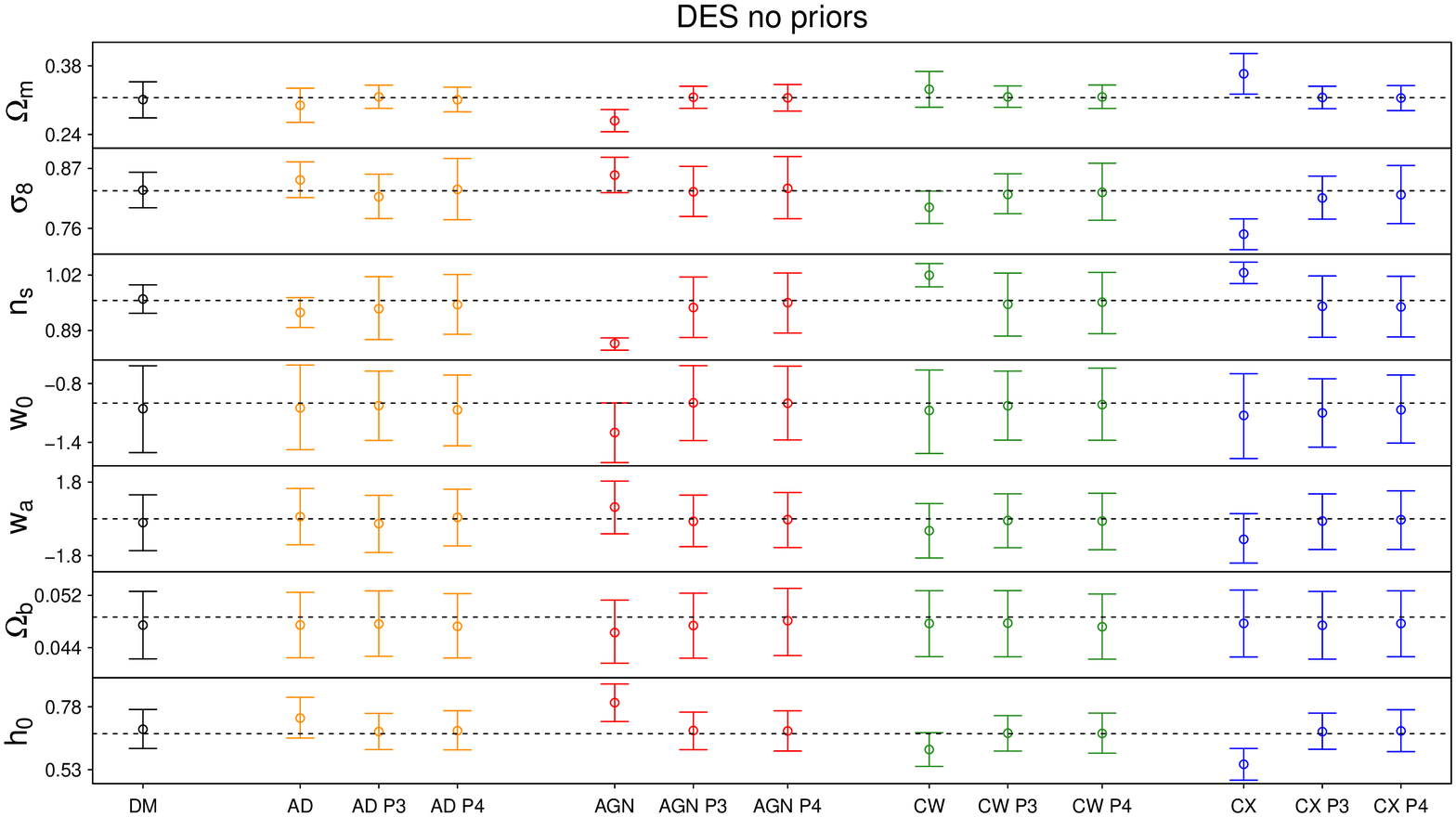}
\caption{The marginalized 1D constraints on cosmological parameters for an DES like survey without priors (see Tab. \ref{tab:paradist_DESnp} for exact numbers). The notation refers to the various simulation scenarios (DM, AD, AGN, CW, CX) and the number of principal components that have been removed from the data, either ``P3" for removal of the three most significant modes or ``P4" for removal of the four most significant modes. }
         \label{fi:errorbias_desnp}
\end{figure*}

\begin{figure*}
\includegraphics[width=18cm]{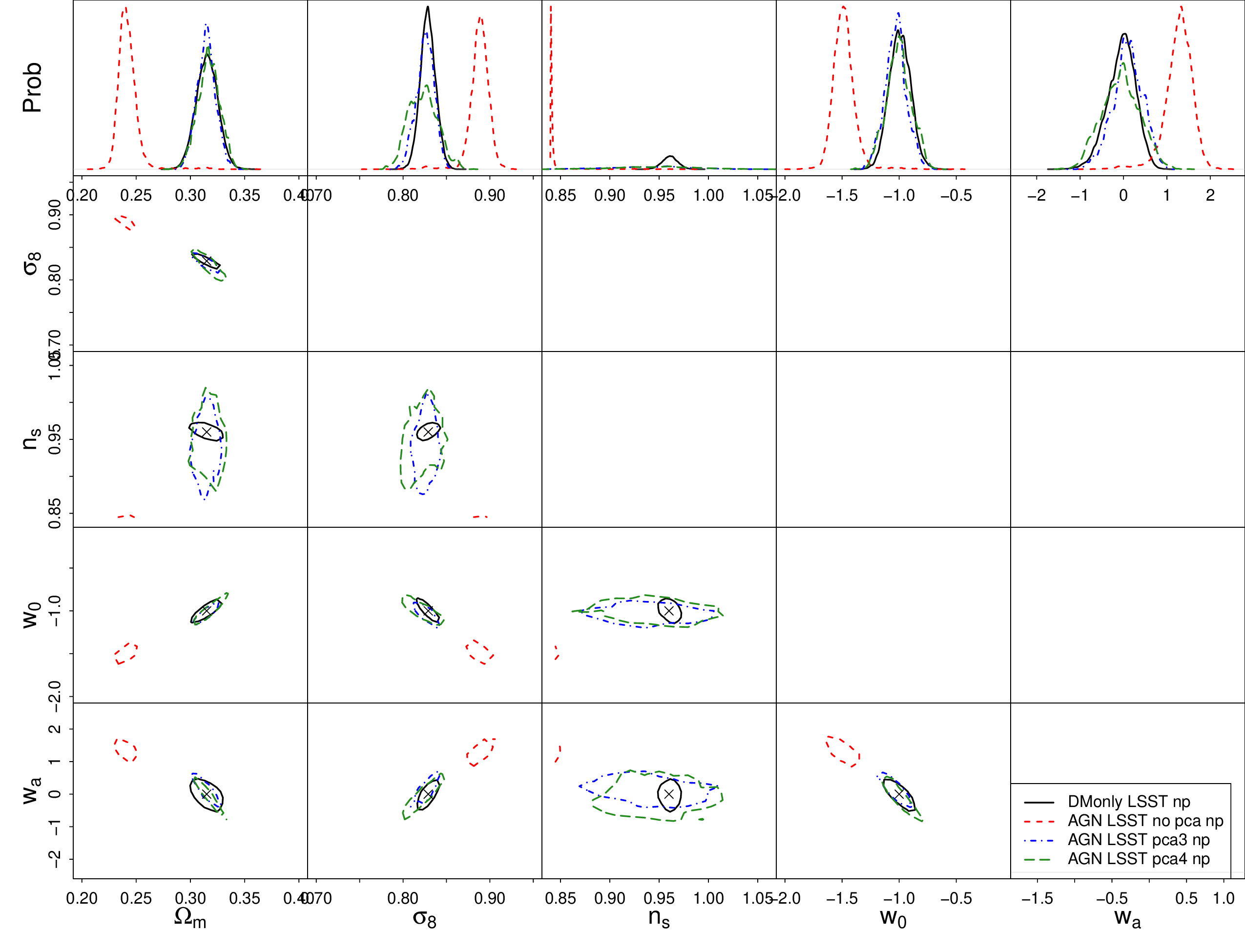}
\caption{Cosmological constraints for a LSST/Euclid survey when using the PCA mitigation technique. The results shown assume that the baryonic physics of the Universe follows the AGN scenario (i.e. the most extreme baryonic scenario). We remove three and four PC modes (\ti{blue/dashed} and \ti{green/long-dashed}, respectively) and compare the results to the untreated AGN scenario (\ti{red/dashed}) and to a pure DM scenario (\ti{black/solid}).}
         \label{fi:lsst_pca_np}
\end{figure*}

\begin{table*}
\caption{Marginalized 1D constraints on cosmological parameters for the DM, AD, AGN, CW, and CX scenario with and without the PCA mitigation for a LSST/Euclid survey (no priors). The last column contains the $\Delta \chi^2$ distance (see Eq. \ref{eq:paradistance}) between best fit and fiducial parameter point.}
\begin{center}
\def\arraystretch{1.3}
\begin{tabular}{|c c c c c c c c c c|}
\hline
Scenario & PCA order & $\om$ & $\sig$ & $\ns$ & $\w$ &$\wa$ & $\omb$ & $h_0$ & $\Delta \chi^2$ \\
\hline
DM&0&0.315$_{-0.00996}^{+0.00982}$ &0.829$_{-0.0087}^{+0.00894}$ &0.961$_{-0.0078}^{+0.00784}$ &-0.994$_{-0.0943}^{+0.0936}$
&-0.0456$_{-0.326}^{+0.318}$
&0.0475$_{-0.00484}^{+0.00483}$
&0.668$_{-0.0276}^{+0.0274}$
&0.33  \\ 
&&&&&&&&&\\ 
AD&0&0.29$_{-0.00988}^{+0.00998}$ &0.857$_{-0.00939}^{+0.00959}$ &0.931$_{-0.00863}^{+0.00816}$ &-1.18$_{-0.0951}^{+0.0951}$
&0.7$_{-0.26}^{+0.27}$
&0.0473$_{-0.00494}^{+0.005}$
&0.742$_{-0.038}^{+0.0378}$
&55.8  \\ 
AD&3&0.319$_{-0.00853}^{+0.00989}$ &0.818$_{-0.0116}^{+0.0117}$ &0.937$_{-0.0521}^{+0.0473}$ &-0.948$_{-0.101}^{+0.112}$
&-0.247$_{-0.434}^{+0.396}$
&0.0476$_{-0.00482}^{+0.0047}$
&0.679$_{-0.0454}^{+0.0454}$
&1.58  \\ 
AD&4&0.315$_{-0.0128}^{+0.0119}$ &0.826$_{-0.0196}^{+0.0204}$ &0.936$_{-0.053}^{+0.0474}$ &-0.992$_{-0.14}^{+0.13}$
&-0.0335$_{-0.503}^{+0.555}$
&0.0473$_{-0.00461}^{+0.00454}$
&0.689$_{-0.0466}^{+0.0504}$
&2.27  \\ 
&&&&&&&&&\\ 
AGN&0&0.242$_{-0.00778}^{+0.00658}$ &0.888$_{-0.00865}^{+0.00995}$ &0.846$_{-0.00608}^{+0.00331}$ &-1.48$_{-0.0854}^{+0.0782}$
&1.27$_{-0.264}^{+0.29}$
&0.042$_{-0.00163}^{+0.00165}$
&0.852$_{-0.0264}^{+0.0347}$
&142  \\ 
AGN&3&0.315$_{-0.00887}^{+0.00858}$ &0.825$_{-0.0105}^{+0.0103}$ &0.939$_{-0.0432}^{+0.0411}$ &-1.03$_{-0.094}^{+0.0927}$
&0.0973$_{-0.349}^{+0.362}$
&0.0471$_{-0.00484}^{+0.00476}$
&0.686$_{-0.0397}^{+0.042}$
&3.55  \\ 
AGN&4&0.317$_{-0.0104}^{+0.0103}$ &0.823$_{-0.0173}^{+0.017}$ &0.947$_{-0.0411}^{+0.041}$ &-1$_{-0.107}^{+0.111}$
&-0.0593$_{-0.446}^{+0.449}$
&0.0483$_{-0.0045}^{+0.00425}$
&0.681$_{-0.0408}^{+0.0408}$
&4.85  \\ 
&&&&&&&&&\\ 
CW&0&0.364$_{-0.0116}^{+0.0126}$ &0.78$_{-0.00956}^{+0.00864}$ &1.02$_{-0.00794}^{+0.0092}$ &-0.597$_{-0.12}^{+0.125}$
&-1.64$_{-0.481}^{+0.429}$
&0.0474$_{-0.00502}^{+0.00495}$
&0.552$_{-0.0243}^{+0.0201}$
&71.5  \\ 
CW&3&0.315$_{-0.00924}^{+0.00958}$ &0.828$_{-0.0115}^{+0.0116}$ &0.947$_{-0.045}^{+0.0429}$ &-0.989$_{-0.1}^{+0.105}$
&0.00541$_{-0.422}^{+0.401}$
&0.0476$_{-0.00483}^{+0.0047}$
&0.679$_{-0.0411}^{+0.0457}$
&0.219  \\ 
CW&4&0.315$_{-0.00952}^{+0.00977}$ &0.83$_{-0.0174}^{+0.017}$ &0.962$_{-0.0506}^{+0.0502}$ &-0.989$_{-0.103}^{+0.105}$
&-0.0353$_{-0.443}^{+0.42}$
&0.0478$_{-0.0048}^{+0.00463}$
&0.668$_{-0.0385}^{+0.0403}$
&0.651  \\ 
&&&&&&&&&\\ 
CX&0&0.431$_{-0.00892}^{+0.0159}$ &0.724$_{-0.00891}^{+0.00385}$ &1.01$_{-0.00765}^{+0.00878}$ &-0.316$_{-0.0458}^{+0.102}$
&-2.3$_{-0.182}^{+0.0701}$
&0.0473$_{-0.00491}^{+0.0049}$
&0.472$_{-0.024}^{+0.0112}$
&86.9  \\ 
CX&3&0.318$_{-0.01}^{+0.00995}$ &0.818$_{-0.0119}^{+0.0121}$ &0.931$_{-0.0447}^{+0.0405}$ &-0.974$_{-0.116}^{+0.116}$
&-0.215$_{-0.489}^{+0.472}$
&0.0483$_{-0.00492}^{+0.00461}$
&0.692$_{-0.0448}^{+0.0492}$
&3.69  \\ 
CX&4&0.321$_{-0.0103}^{+0.0104}$ &0.812$_{-0.0155}^{+0.0154}$ &0.917$_{-0.0459}^{+0.0436}$ &-0.936$_{-0.118}^{+0.117}$
&-0.322$_{-0.453}^{+0.465}$
&0.0477$_{-0.00468}^{+0.00448}$
&0.692$_{-0.0419}^{+0.0416}$
&4.11  \\ 
\hline
\end{tabular}
\end{center}
\label{tab:paradist_LSSTnp}
\end{table*}

\begin{figure*}
\includegraphics[width=18cm]{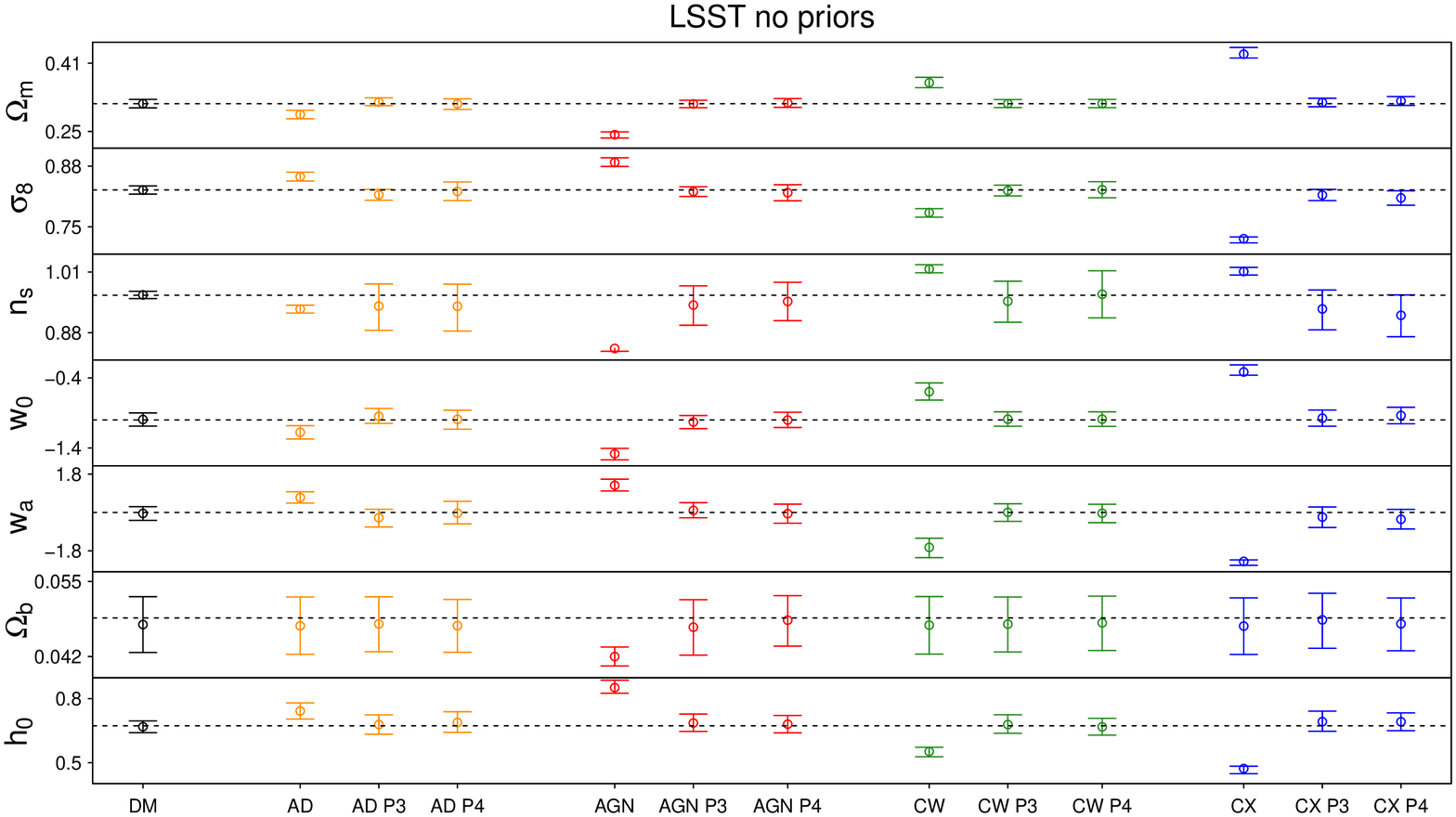}
\caption{The marginalized 1D constraints on cosmological parameters for an LSST like survey without priors (see Tab. \ref{tab:paradist_LSSTnp} for exact numbers). The notation refers to the various simulation scenarios (DM, AD, AGN, CW, CX) and the number of principal components that have been removed from the data, either ``P3" for removal of the three most significant modes or ``P4" for removal of the four most significant modes.}
         \label{fi:errorbias_lsstnp}
\end{figure*}

From Fig. \ref{fi:powspec} and Table \ref{tab:barypcasub} we expect that removing three or four PCs is sufficient to remove any bias from baryonic physics. We now test this by running the likelihood analyses with differing numbers of modes removed. 

It is important to note that all likelihood analyses with PCA marginalization are blind to the baryonic scenario that serves as input data vector. More precisely, this means that we {\em exclude the data vector's baryonic scenario from the matrix $\Del$} in order not to have an unfair advantage over reality.

Figures \ref{fi:des_pca_np} and \ref{fi:lsst_pca_np} show the results of the likelihood analyses after removing zero (\ti{red, dashed}), three (\ti{blue, dot-dashed}), and four (\ti{green, long-dashed}) modes for a DES and LSST/Euclid like survey, respectively. In comparison we show a pure DM input scenario analyzed with a DM prediction code (\ti{black, solid}). All contour plots are marginalized over five cosmological parameters; in addition to the ones mentioned in the plots we marginalize over $\omb$ and $H_0$. The first row of Figs. \ref{fi:des_pca_np} and \ref{fi:lsst_pca_np} show the posterior probability distribution of a given cosmological parameter marginalized over the other 6 cosmological parameters. In each case, the ``data'' is taken to be the spectra from the AGN simulation, which -- as depicted in Figs.~\ref{fi:des_np} and \ref{fi:lsst_np} and by the red dashed curves -- led to the largest biases if baryons were not accounted for. The LSST/Euclid plot shows that, after removing 3 or 4 modes, the bias vanishes. 

One typically expects that a mitigation scheme that removes the bias will loosen constraints (e.g., adding extra nuisance parameters to capture the effects of the systematic will inevitably degrade the marginalized constraints on the cosmological parameters). In our mitigation scheme, we are removing some of the data so we similarly expect some degradation in the constraints. Figures \ref{fi:des_pca_np} and \ref{fi:lsst_pca_np} show that this degradation is minimal, affecting only the spectral index $n_s$. Again, this is an exciting result: the mitigation scheme can be used with little cost to the overall extraction. It is perhaps not surprising that the one parameter that is affected, $n_s$, is the one that requires information from both large and small scales. By removing some of the small scale information, we are necessarily losing information about $n_s$.

\begin{table*}
\caption{Projection angle of the dark matter difference vectors onto the PCs (see Eq. \ref{eq:angle}) and fraction of the difference vector contained by the PC subspace (see Eq. \ref{eq:vol}). Results shown are for the DES case.}
\begin{center}
\def\arraystretch{1.3}
\begin{tabular}{|l|r c|r c|r c|r c|}
\hline
Cosmology & $\abs{\cos \theta_1}$ & $V_1$ & $\abs{\cos \theta_2}$ & $V_2$& $\abs{\cos \theta_3}$ & $V_3$ &$\abs{\cos \theta_4}$ & $V_4$ \\ \hline 

$\om=$0.313$_{-0.0395}^{+0.0341}$ & 0.32 & 0.32 & 0.45 & 0.55 & 0.017 & 0.55 & 0.22 & 0.6  \\
$\sig=$0.83$_{-0.0318}^{+0.0335}$ & 0.18 & 0.18 & 0.49 & 0.52 & 0.14 & 0.54 & 0.2 & 0.58  \\
$\ns=$0.964$_{-0.0337}^{+0.033}$ & 0.26 & 0.26 & 0.55 & 0.61 & 0.5 & 0.79 & 0.044 & 0.79  \\
$\w=$-0.972$_{-0.532}^{+0.353}$ & 0.14 & 0.14 & 0.61 & 0.62 & 0.28 & 0.68 & 0.14 & 0.7  \\
$\wa=$0.372$_{-1.18}^{+1.55}$  & 0.22 & 0.22 & 0.61 & 0.65 & 0.3 & 0.72 & 0.14 & 0.73  \\
$\omb=$0.041$_{-0.00131}^{+0.0116}$ & 0.36 & 0.36 & 0.38 & 0.53 & 0.52 & 0.74 & 0.029 & 0.75  \\
$h_0=$0.672$_{-0.0575}^{+0.0974}$ & 0.37 & 0.37 & 0.41 & 0.55 & 0.52 & 0.76 & 0.043 & 0.76  \\
\hline
\end{tabular}
\end{center}
\label{tab:cosmopcasub_DES}
\end{table*}

\begin{table*}
\caption{Projection angle of the dark matter difference vectors onto the PCs (see Eq. \ref{eq:angle}) and fraction of the difference vector contained by the PC subspace (see Eq. \ref{eq:vol}). Results shown are for the LSST/Euclid case.}
\begin{center}
\def\arraystretch{1.3}
\begin{tabular}{|l|r c|r c|r c|r c|}
\hline
Cosmology & $\abs{\cos \theta_1}$ & $V_1$ & $\abs{\cos \theta_2}$ & $V_2$& $\abs{\cos \theta_3}$ & $V_3$ &$\abs{\cos \theta_4}$ & $V_4$ \\ \hline
$\om=$0.315$_{-0.00961}^{+0.0102}$ & 0.099 & 0.099 & 0.59 & 0.59 & 0.32 & 0.67 & 0.13 & 0.69  \\
$\sig=$ 0.829$_{-0.00862}^{+0.00901}$ & 0.15 & 0.15 & 0.49 & 0.51 & 0.15 & 0.53 & 0.28 & 0.6  \\
$\ns=$0.961$_{-0.00793}^{+0.00771}$ & 0.13 & 0.13 & 0.063 & 0.15 & 0.15 & 0.21 & 0.4 & 0.45  \\
$\w=$-1.01$_{-0.0776}^{+0.11}$ & 0.14 & 0.14 & 0.37 & 0.4 & 0.081 & 0.41 & 0.32 & 0.52  \\
$\wa=$0.0402$_{-0.412}^{+0.232}$ & 0.14 & 0.14 & 0.44 & 0.46 & 0.12 & 0.48 & 0.19 & 0.52  \\
$\omb=$0.0486$_{-0.00593}^{+0.00374}$ & 0.13 & 0.13 & 0.054 & 0.14 & 0.26 & 0.3 & 0.43 & 0.52  \\
$h_0=$0.673$_{-0.0319}^{+0.0232}$ & 0.13 & 0.13 & 0.038 & 0.14 & 0.24 & 0.28 & 0.43 & 0.51  \\
\hline
\end{tabular}
\end{center}
\label{tab:cosmopcasub_LSST}
\end{table*}

We can quantify the extent to which the bias is removed and the amount by which the allowed region in parameter space is broadened by the mitigation scheme. If there were only one parameter, this would be straightforward: simply report the difference between the best value of the parameter emerging from the likelihood analysis and the ``true'' value used to generate the spectra. This would be the bias, and it would be compared to the statistical uncertainty emerging from the likelihood analysis. Bias significantly smaller than this uncertainty would be fine, while one larger would be a problem.
That is, the relevant quantity would be $(p^{\rm best\, fit} - p^{\rm fid})^2/\sigma^2$. Under the assumption that this $\Delta\chi^2$ is drawn from a chi squared distribution, a value larger than one would indicate a problem at 68\%; larger than 4 at 95\%; and larger than 9 at 99.7\%.

For our seven parameter case, we generalize to 
\be
\label{eq:paradistance}
\Delta \chi^2 = (\pco^\mr{fid} - \pco^\mr{bary, best \, fit})^t \, \matC_{\pco}^{-1} \, (\pco^\mr{fid} - \pco^\mr{bary, best\, fit})
,\ee
where the covariance matrix is determined via
\be
\label{eq:paracov}
\matC_{\pco}^{ij} = \frac{1}{N-1} \sum_{k=0}^N \left(\langle \pco^i \rangle - \pco^{ik} \right) \left(\langle \pco^j \rangle - \pco^{jk}\right)\,
\ee
with $\langle \pco^i \rangle$ indicating the mean of the $i$-th cosmological parameter ($i,j \in [1,7]$), and $k \in [1,N]$ being the index running over all steps in the MCMC chain. Again assuming this is distributed in the seven-dimensional cosmological parameter space as a $\chi^2$ distribution with seven degrees of freedom, we find the critical $\Delta \chi^2$ values that correspond to $68\%$, $95\%$, and $99\%$ confidence regions are 8.14, 14.07, and 18.48, respectively. 

In Tables \ref{tab:paradist_DESnp} and \ref{tab:paradist_LSSTnp} (also see Figs. \ref{fi:errorbias_desnp} and \ref{fi:errorbias_lsstnp} ), we show the best fit values of the individual parameters with the marginalized error bars and the $\Delta \chi^2$ as defined in Eq. (\ref{eq:paradistance}). This analysis illustrates the severe biases in cosmological constraints for DES if the extreme baryonic scenarios are analyzed. For example, when analyzing the AGN feedback scenario the probability of the fiducial cosmology is outside the $\alpha=99.9999998\%$ confidence interval. For scenarios that only slightly differ from a pure DM Universe, such as the adiabatic (AD) scenario the bias is substantially less severe (within the $68\%$ region) but still noteworthy.

As expected the impact of baryonic physics is more important for Stage IV surveys. For example, the analysis of the AD scenario for an LSST/Euclid-like experiment rejects the fiducial cosmology more strongly (outside the $\alpha=99.9999999\%$ confidence interval) than the AGN scenario does for DES. When analyzing the AGN scenario for a LSST/Euclid survey, the fiducial cosmology is outside the $\alpha=\exp(-5 \times 10^{-27})$ (a number that is considered 1 by almost any calculator) interval. 

Focusing on the LSST/Euclid case, we see that - in accord with the 2D projections shown in the figures - the biases are extremely large for all baryonic scenarios if no mitigation scheme is used. As more modes are removed, the fits get significantly better, e.g.,  $\Delta \chi^2$ drops from 55.8 to 1.58 and 2.27 for the AD scenario when removing 3 and 4 PCs, respectively. For the AGN scenario we find a similar behavior for $\Delta \chi^2$, i.e. it drops from 142 to 3.55, 4.85 when removing 3 and 4 PCs, respectively. For all considered scenarios the bias is well within the 1-$\sigma$ error bars, hence we conclude that the mitigation scheme effectively removes the baryonic bias even for Stage IV surveys such as LSST and Euclid. This is in distinct contrast to phenomenological models, such as those studied in \citet{zsd13} and \citet{shs13}, which are adequate for Stage III surveys such as DES, but leave significant systematic error in the inferred cosmological parameters from Stage IV experiments.

In Appendix~\ref{sec:app}, we rerun all likelihood analyses described in this section and in Sect. \ref{sec:like_without} but include prior information from the Planck mission. Figures to compare are Figs. \ref{fi:des_np} and \ref{fi:lsst_np} to Figs. \ref{fi:des_pp} and \ref{fi:lsst_pp} for the impact of baryonic physics on constraints without any mitigation and Figs. \ref{fi:des_pca_np} and \ref{fi:lsst_pca_np} to Figs. \ref{fi:des_pca_pp} and \ref{fi:lsst_pca_pp} for the likelihood analyses with PCA marginalization. We also repeat the analyses of Tables \ref{tab:paradist_DESnp}, \ref{tab:paradist_LSSTnp} and Figs \ref{fi:errorbias_desnp}, \ref{fi:errorbias_lsstnp}, which are mirrored in Tables \ref{tab:paradist_DESpp}, \ref{tab:paradist_LSSTpp} and Figs. \ref{fi:errorbias_despp}, \ref{fi:errorbias_lsstpp}, respectively. 

The inclusion of Planck information (which in our implementation does not act on $w_0$, $w_a$) mitigates the magnitude of the bias from the cosmic shear tomography analysis; however, it also substantially reduces the statistical errors on cosmological parameters, and this places stronger demands on the performance of any mitigation scheme. Qualitatively, the results with and without Planck information are similar: First, we find significant biases in cosmological constraints if baryonic physics is not accounted for; the biases are severe for DES and catastrophic for LSST/Euclid. Second, PCA marginalization is able to remove these biases efficiently. One major difference between both analyses is that the information loss on $n_s$ is insignificant when including Planck information. In this case, the Planck prior determines the constraint on $n_s$ entirely.

\subsection{Degeneracy with cosmological parameters}
\label{sec:pcasub}

As shown in Figs. \ref{fi:des_pca_np}-\ref{fi:errorbias_lsstnp} the PCA removal technique substantially reduces the information on the spectral index $n_s$ indicating a strong degeneracy of baryonic scenarios and this particular cosmological parameter. In order to investigate this degeneracy further we perform a similar analysis as in Table \ref{tab:barypcasub} but replacing the $\vek \Delta_\alpha$ in Eqs. (\ref{eq:angle}, \ref{eq:vol}) with the difference of dark matter data vectors that vary in their underlying cosmology (see Tables \ref{tab:cosmopcasub_DES} and \ref{tab:cosmopcasub_LSST}). 

Specifically, we compute the difference vectors between the DM fiducial model and the $68\%$ intervals for each of the seven cosmological parameters considered in our likelihood analysis. A second difference to the analysis in Table \ref{tab:barypcasub} is the inclusion of the covariance matrix of the $C^{ij}(l)$ when deriving the PCs. As we will further outline in Sect. \ref{sec:general} (see Eq. \ref{eq:set2}) this accounts for correlation and different error bars on the individual $C^{ij}(l)$.

A sufficient but not a necessary condition for the removal of the information on $n_s$ would be $V_4 \sim 1$, which however is not reflected in Tables \ref{tab:cosmopcasub_DES} (DES) and \ref{tab:cosmopcasub_LSST} (LSST/Euclid). Whereas for the DES case one might argue that $V_4$ of $n_s$ has the largest value of all cosmological parameters the other values are too close to draw any conclusions from this analysis. Especially since for the LSST/Euclid analysis we find that the volume of $\om$ and $\sig$ is more prominently mapped onto the PC-space than $n_s$. 

Our explanation for this is that $n_s$ is only degenerate with the PCs if additional cosmological parameters are at least allowed to vary slightly as well. We motivate this statement as follows: Suppose $n_s$ were the only parameter of interest. Under variation of $n_s$ the power spectrum gets tilted, hence the difference vector has contributions from small and large scales. However, baryonic scenarios only act on small scales, hence when all other parameters, the removal of baryons will not void the information on $n_S$. Given some freedom in especially $\om$, $\sig$, and $\w$, the spectral index $n_s$ can indeed account for the tilts that are seen in most baryonic scenarios. We have examined some combinations of the aforementioned parameters, finding indeed that $V_4$ for $n_s$ strongly increases already when giving only little freedom to $\sig$ and $\om$. We however postpone a more thorough study of the cosmological parameter space degeneracies to a future paper.

\section{Generality and discussion of the method}
\label{sec:general}

The PCA mitigation technique introduced in Sect. \ref{sec:likebasics} is completely general and can be applied to any quantity that enters a likelihood analysis and any (combinations of) systematic(s) that affect said quantity. In this section we formalize and discuss a general PCA marginalization scheme; the main differences to the method outlined in Sect. \ref{sec:likebasics} are that we require the method to be (i) agnostic about the DM scenario (ii) account for multiple systematics, (iii) account for correlation and different errors of observables, and (iv) to be able to process prior information on a systematics scenario (e.g., the AGN scenario being more likely to resemble the true baryonic physics compare to the AD scenario).

The first requirement is motivated by the fact that even if one can reference to a DM power spectrum, non-linear density evolution models of the DM power spectrum itself are affected by uncertainties that need to be marginalized over. For example, even the latest Coyote Universe emulator \cite{hlk14} has up to $5\%$ uncertainties in the DM power spectrum and \cite{eif11} showed that this can substantially impact weak lensing observables. The \textsc{CosmoLike} weak lensing module employed in this paper \citep[i.e.,][with a modification to include time-dependent dark energy models]{tsn12} is likely to exceed the 5\% uncertainty threshold at small $k$-modes. This uncertainty should be accounted for, hence we conclude that referencing to the (weighted) mean of all models is a more objective choice.

Consequentially, we define the components of the difference matrix not with respect to a DM scenario (as in Eq. \ref{eq:diffmat}) but to the mean of all models
\be
\label{eq:delta}
\Delta_{k\alpha}=M_{k\alpha} - \bar M_k \qquad \mr{with} \qquad \bar M_k=\frac{1}{N_\mr{sce}-1} \sum^\mr{N_{sce}}_{\alpha} M_{k\alpha}\,.
\ee
where $k$ again labels the model vector bin in $(l, z)$, and $\alpha$ refers to the various systematic scenarios. 

The difference matrix is again computed at every point of the MCMC and the $\M_\alpha$'s resemble uncertainties from systematics at any given point in cosmology, i.e. $\M(\pnu| \pco)$. In order to account for requirement (iii) and (iv) we have to modify the $\M(\pnu| \pco)$ and define the $\M_\alpha$'s as
\be
\label{eq:set2}
\M_\alpha = w_\alpha \; \mathbf{L} \; \M(\pnu| \pco)  \,
\ee
where the $w_\alpha$'s allow the analyst to weigh the different nuisance parameter scenarios relative to each other and $\mathbf L$ is computed from the inverse data covariance matrix $\mathbf C^{-1}=\mathbf L \mathbf L^t$ in order to account for correlation and different error bars of data points. We note that strictly speaking the covariance is a function of $\pnu$ and $\pco$ and that this dependency should be incorporated in a high precision analysis. 

In order to fulfill requirement (ii), it must be possible to compute the effect of the systematic under a wide range of possible circumstances. This computation involves information from observations, simulations and theoretical considerations; it is necessary for our calculations to span the range of reasonable realizations of the systematic effect. PCA mitigation does not eliminate the need to produce simulations of the systematics that one aims to remove. The procedure also requires that the systematic not be largely degenerate with the parameters we aim to infer from the data; however, this same requirement must be met for more commonplace ``self-calibration" exercises to be effective \citep[e.g., such as][]{htb06,zrh08,ber09,shs13,zsd13}. 

There are substantial advantages of this technique over other nuisance parameter approaches. First and foremost, the process is bound to effectively incorporate degeneracies between models of systematic uncertainties. This is not true if independently developed nuisance parameter models, e.g., baryons as in \cite{zsd13} and intrinsic alignment as in \cite{jma11} are combined in an analysis. Second, if systematics can be calibrated against dark matter only simulation, this procedure enables one to perform a cosmological analysis using phenomenological models that require relatively little computational effort. This advantage should not be underestimated. The computational expense of e.g., explicitly including baryonic effects in simulations for a wide range of cosmological models, is so prohibitive as to be entirely infeasible. Third, the technique to remove contaminated modes substantially reduces the dimensionality of the parameter space that needs to be sampled. Instead of sampling a high-dimensional nuisance parameter space at every step of the MCMC, mode removal allows the analyst to sample cosmological parameters only.

In the presence of strong degeneracies between PCs and cosmology the mode removal technique might need to be replaced by marginalizing over the PCs with priors (recall that mode removal is equivalent to marginalizing without priors). This changes the formalism outlined in Sect. \ref{sec:likebasics}. Instead of removing the contaminated modes as in Eq. (\ref{eq:chipc2}) we have to carry out a full marginalization in PC space. 

Defining data and model vector and covariance in the nuisance parameter sensitive PC space, i.e. $\D_\mr{pc}=\vek U^t \D$, $\M_\mr{pc}=\vek U^t \M$, and $\matC^{-1}_\mr{pc}=\vek U^t  \matC^{-1} \vek U$, we can define the marginalization integral that needs to be solved/computed at every step of the MCMC as
\bea
\label{eq:likemarg2}
L (\D|\pco ) &=&   \int  \d^n \mr{pc}^i \, Pr(\mr{pc^i}) \\ \nn
&\qquad& \times  \exp \biggl( -\frac{1}{2}  \left[ (\D_\mr{pc} - \M_\mr{pc})^t \matC^{-1}_\mr{pc} (\D_\mr{pc} - \M_\mr{pc}) \right]  \biggr)  \,,
\eea
where $Pr(\mr{pc^i})$ accounts for prior information on the $i$-th PC. Such information can be obtained from the eigenvalues of the covariance matrix in Eq. (\ref{eq:singtocov}) or from the so-called \ti{signals} of the extreme $\M_\alpha$'s, i.e. their projection onto the PC's. These extreme signals can serve as upper and lower integration limit of the marginalization integral. 

We note however that even in this scenario the PC mitigation technique has substantial advantages: (i) the degeneracy between nuisance parameters is automatically accounted for and (ii) the number of nuisance parameters and hence the dimensionality of the integral is greatly reduced. 

\section{Conclusions}
\label{sec:conc}
We analyze cosmic shear tomography power spectra obtained from 14 hydro-simulations with different underlying baryonic processes (e.g., AGN feedback, SN feedback, different cooling mechanisms, and combinations thereof). These simulations span the range of modeling uncertainties in the matter density field which, if not accounted for, severely impact cosmological constraints.

Using the covariance and weak lensing module of the \textsc{CosmoLike} analysis framework, we simulate Stage III (DES) and Stage IV (LSST/Euclid) likelihood analyses for each of the 14 scenarios. The quantity of interest is the bias in the inferred parameter (e.g., $w_0$, $w_a$, $\sigma_8$) caused by baryonic effects compared to the statistical uncertainties in the inferred parameter. In agreement with previous, similar analyses \citep[e.g.,][]{shs11,shs13, zsd13}, we find severe biases in cosmological constraints inferred from cosmic shear measurements of DES {\em if} the true Universe is described by one of the extreme baryonic scenarios {\em and} baryonic effects are neglected in the analysis. For scenarios that differ only slightly from a pure DM Universe, such as the adiabatic (AD) scenario the bias is substantially less severe (within the $68\%$ region) but still non-negligible. Unfortunately, detailed studies of the OWLS simulations analyzed here suggest that some of the more extreme scenarios best describe observed galaxy properties \citep[e.g.,][]{msp10}. 

The Stage IV experiments LSST and Euclid will measure cosmic shear spectra with smaller statistical error bars and so the requirement to reduce systematics is significantly more stringent than for DES. In our analyses in which we use baryonic simulations to simulate an observed LSST/Euclid cosmic shear data set, but do not account for baryonic effects, the systematic errors on inferred cosmological parameters are always severe. In these analyses, biases in dark energy equation of state parameters could be as large as $\sim 7\sigma$, while biases in other parameters could be even larger. In the AD scenario, in which the baryons are treated non-radiatively and in which the alterations of cosmic shear spectra are mild, our analysis that does not include specific mitigation of baryonic effects rejected the true, fiducial cosmology at greater than the $\alpha=99.9999999\%$ confidence interval. When analyzing the AGN scenario for an LSST/Euclid survey the rejection probability of the fiducial cosmology is off the charts. We repeat all likelihood analyses including prior information from the Planck mission and find no qualitative change in the severity of the effect (see Appendix~\ref{sec:app}). There is no doubt then that a mitigation scheme will be necessary to analyze both Stage III and certainly Stage IV data. 

As a potential remedy we present PCA marginalization which aims to mitigate biases on parameters inferred from observables that may be partly compromised by poorly-understood systematic errors. The technique consists of: (i) identifying a range of possible effects that the systematic may have on the observable of interest; (ii) determining linear combinations of observables, using a principal component analysis, that are most compromised by the systematic according to the templates identified in step (i); projecting the data onto a subspace that removes the linear combinations of observables that are most affected by the systematic; and (iii) performing a likelihood analysis on this data subspace. 

We apply PCA marginalization to the simulation data and repeat the likelihood analyses for all baryonic scenarios. We find that removing 3-4 principal components is sufficient to account fully for biases from baryonic physics, even for the most extreme baryonic scenarios, and even for the Stage IV LSST/Euclid surveys. This is a clear improvement over phenomenological models \citep[as in][]{zsd13,shs13}, which remove biases from baryonic physics for Stage III, but leave significant systematic error in the inferred cosmological parameters from Stage IV experiments.

As a consequence of the PCA mode removal, our constraining power on cosmology is only slightly reduced, with only the constraints on the spectral index $n_s$ noticeably degraded. Even this loss in cosmological information is recaptured if Planck priors are included. Accounting for both the statistical and systematic errors on inferred cosmological parameters (such as $w_0$ and $w_a$), it is clear that mitigation is strongly preferred over neglecting baryonic processes. For example, in the LSST/Euclid AGN scenario in which the baryonic systematic is not explicitly mitigated, the systematic error on $w_0$ is $\delta w_0 \approx 0.5$, while the statistical error is $\sigma(w_0) \approx 0.08$ (see Table~\ref{tab:paradist_LSSTnp}). Upon applying PCA mitigation to this scenario and removing the three most important modes, the systematic error is reduced to $\delta w_0 \approx 0.03$ and the statistical error increased to only $\sigma(w_0) \approx 0.09$.  

It is our hope that these techniques will be adopted and applied to mitigate systematic errors, not only in cosmic shear cosmology, but in a variety of future data analyses.

\section*{Acknowledgments}

This paper is based upon work supported in part by the National Science Foundation under Grant No. 1066293 and the hospitality of the Aspen Center for Physics. The work of SD and NG is supported by the U.S. Department of Energy, including grant DE-FG02-95ER40896. The work of ARZ has been funded in part by the Pittsburgh Particle physics, Astrophysics, and Cosmology Center (PITT PACC) at the University of Pittsburgh and by the National Science Foundation under grants AST 0806367 and  AST 1108802. Support for this work was provided through the Scientific
Discovery through Advanced Computing (SciDAC) program funded by the U.S. Department of Energy, Office of Science, Advanced Scientific Computing Research and High Energy Physics. Part of the research was carried out at the Jet Propulsion Laboratory, California Institute of Technology, under a contract with the National Aeronautics and Space Administration.

\begin{appendix}

\section{Results with Planck Priors}
\label{sec:app}
In this appendix we repeat all likelihood analysis described in the text but include external information from the Planck satellite \citep{planckcosmo13}. There are good reasons to look at the no-prior likelihoods first. Before combining results, we would need to see whether they are consistent, and -- were baryons neglected -- the Planck results would not be consistent with the lensing results. There is also the danger that including external results would force the likelihood to the correct region, thereby understating the magnitude of the problem and the need to fix it. 

The analysis methods used in this appendix are exactly the same as in the main text; the results are described in Sect. \ref{sec:results2}.

\begin{figure*}
  \begin{minipage}[c]{0.85\textwidth}
\includegraphics[width=\textwidth]{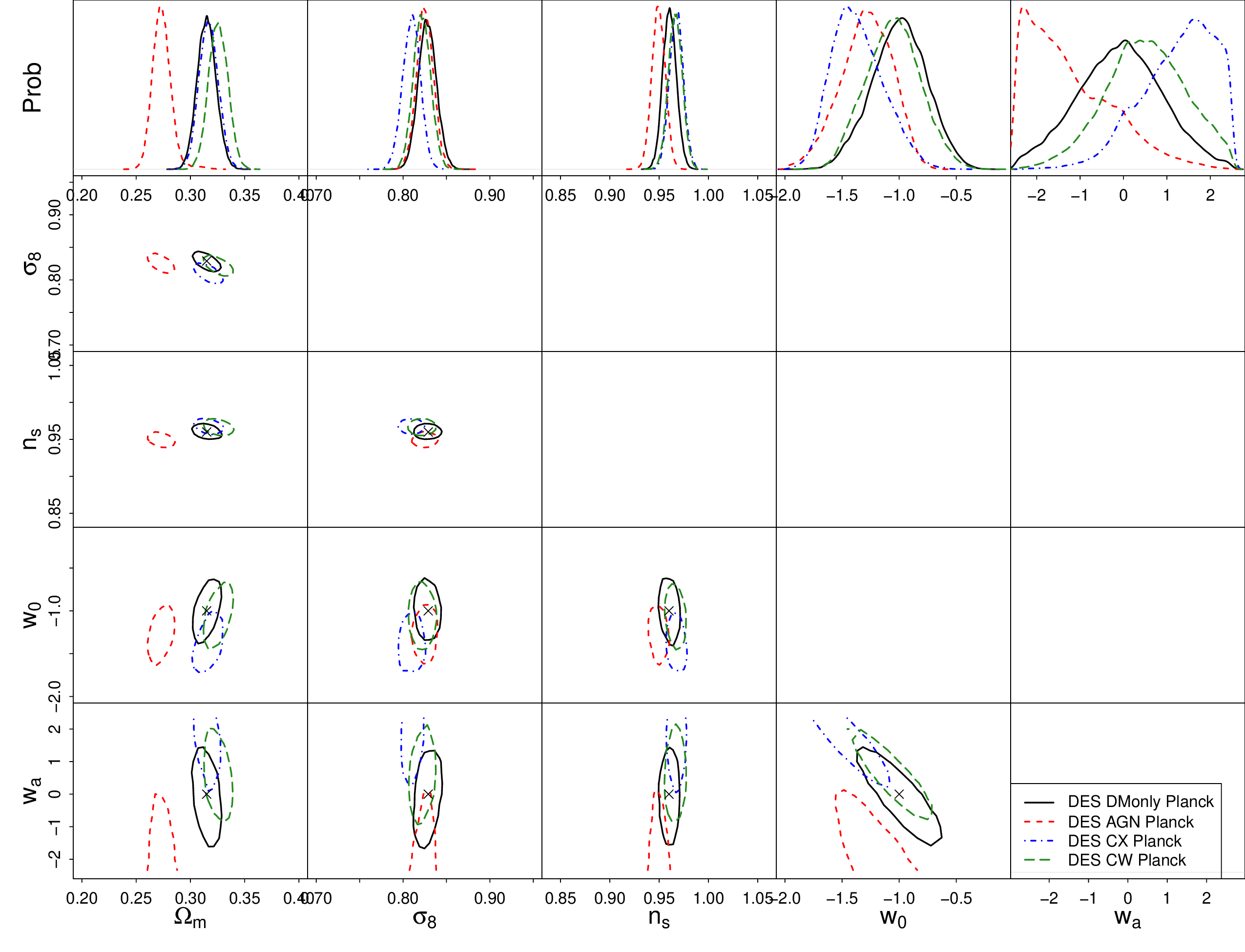}
  \end{minipage}\hfill
  \begin{minipage}[c]{0.13\textwidth}
   \caption{Cosmological constraints for a DES survey assuming different underlying baryonic scenarios for our Universe, i.e. pure dark matter \ti{(black, solid)}, strong AGN feedback \ti{(red, dashed)}, extreme cooling \ti{(blue, dot-dashed)}, and moderate cooling \ti{(green, long-dashed)}. The scenarios are detailed in Sect. \ref{sec:sim}. Results shown here include priors front the Planck mission.} \label{fi:des_pp}
  \end{minipage}
  \begin{minipage}[c]{0.85\textwidth}
\includegraphics[width=\textwidth]{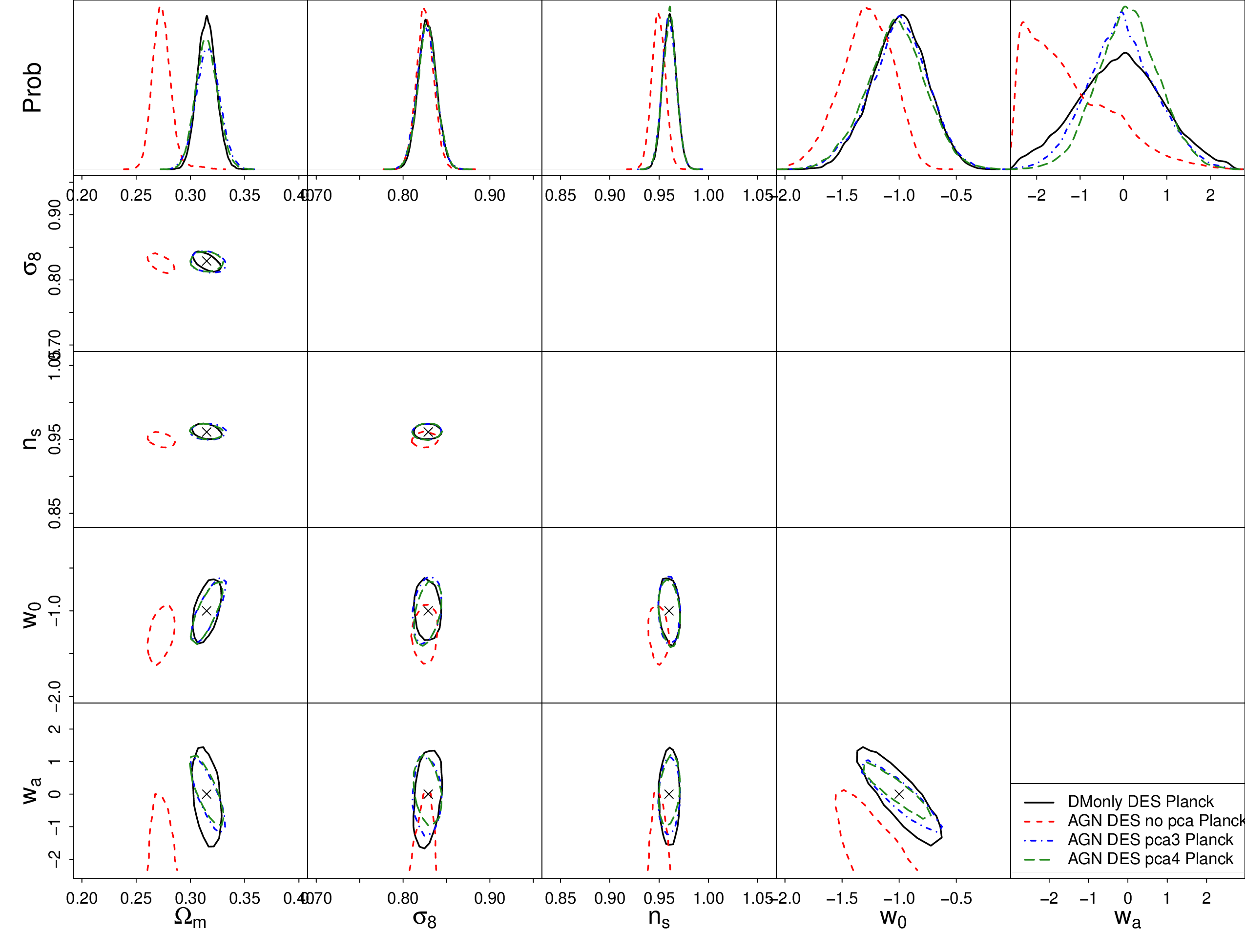}
  \end{minipage}\hfill
  \begin{minipage}[c]{0.13\textwidth}
      \caption{Cosmological constraints for a DES survey with Planck priors when using the PCA mitigation technique. The results shown assume that the baryonic physics of the Universe follows the AGN scenario (i.e. the most extreme baryonic scenario). We remove three and four PC modes (\ti{blue/dashed} and \ti{green/long-dashed}, respectively) and compare the results to the untreated AGN scenario (\ti{red/dashed}) and to a pure DM scenario (\ti{black/solid}).}\label{fi:des_pca_pp}
  \end{minipage}
\end{figure*}
\begin{table*}
\caption{Marginalized 1D constraints on cosmological parameters for the DM, AD, AGN, CW, and CX scenario with and without the PCA mitigation for a DES survey (with Planck priors). The last column contains the $\Delta \chi^2$ distance (see Eq. \ref{eq:paradistance}) between best fit and fiducial parameter point.}
\begin{center}
\def\arraystretch{1.3}
\begin{tabular}{|c c c c c c c c c c|}
\hline
Scenario & PCA order & $\om$ & $\sig$ & $\ns$ & $\w$ &$\wa$ & $\omb$ & $h_0$ &$\Delta \chi^2$ \\
\hline
DM&0&0.315$_{-0.00862}^{+0.00856}$ &0.828$_{-0.0103}^{+0.0104}$ &0.961$_{-0.00686}^{+0.00687}$ &-0.999$_{-0.24}^{+0.24}$
&-0.0836$_{-0.995}^{+0.974}$
&0.0487$_{-0.000607}^{+0.000602}$
&0.673$_{-0.0116}^{+0.0118}$
&0.0428  \\ 
&&&&&&&&&\\ 
AD&0&0.308$_{-0.00819}^{+0.00815}$ &0.83$_{-0.0104}^{+0.0105}$ &0.958$_{-0.00705}^{+0.00695}$ &-0.974$_{-0.223}^{+0.226}$
&-0.528$_{-1.04}^{+0.98}$
&0.0487$_{-0.000616}^{+0.000617}$
&0.673$_{-0.0117}^{+0.0118}$
&2.82  \\ 
AD&3&0.313$_{-0.00886}^{+0.00872}$ &0.827$_{-0.0109}^{+0.0109}$ &0.96$_{-0.00702}^{+0.00712}$ &-1.05$_{-0.221}^{+0.22}$
&0.0795$_{-0.642}^{+0.647}$
&0.0487$_{-0.000635}^{+0.00063}$
&0.672$_{-0.0107}^{+0.0107}$
&0.148  \\ 
AD&4&0.314$_{-0.0107}^{+0.0108}$ &0.828$_{-0.0108}^{+0.0106}$ &0.96$_{-0.0069}^{+0.00687}$ &-1.03$_{-0.251}^{+0.249}$
&0.0335$_{-0.703}^{+0.702}$
&0.0487$_{-0.000629}^{+0.000628}$
&0.673$_{-0.0107}^{+0.0107}$
&0.258  \\ 
&&&&&&&&&\\ 
AGN&0&0.274$_{-0.00844}^{+0.00791}$ &0.825$_{-0.00967}^{+0.00975}$ &0.949$_{-0.00674}^{+0.0068}$ &-1.29$_{-0.228}^{+0.229}$
&-1.22$_{-0.988}^{+1.07}$
&0.0486$_{-0.000623}^{+0.000614}$
&0.671$_{-0.0118}^{+0.012}$
&60.5  \\ 
AGN&3&0.316$_{-0.0108}^{+0.0107}$ &0.827$_{-0.0109}^{+0.0109}$ &0.96$_{-0.00701}^{+0.00712}$ &-1.01$_{-0.258}^{+0.255}$
&-0.0336$_{-0.78}^{+0.794}$
&0.0487$_{-0.00062}^{+0.000623}$
&0.674$_{-0.0112}^{+0.0114}$
&0.797  \\ 
AGN&4&0.315$_{-0.0102}^{+0.0102}$ &0.828$_{-0.0107}^{+0.0107}$ &0.96$_{-0.00708}^{+0.00699}$ &-1.03$_{-0.257}^{+0.255}$
&0.108$_{-0.703}^{+0.708}$
&0.0487$_{-0.000614}^{+0.000611}$
&0.673$_{-0.0105}^{+0.0103}$
&0.538  \\ 
&&&&&&&&&\\ 
CW&0&0.326$_{-0.00874}^{+0.00882}$ &0.822$_{-0.00988}^{+0.00994}$ &0.966$_{-0.00698}^{+0.00702}$ &-1.05$_{-0.243}^{+0.236}$
&0.504$_{-0.908}^{+0.95}$
&0.0487$_{-0.000614}^{+0.000613}$
&0.674$_{-0.0113}^{+0.0114}$
&5.28  \\ 
CW&3&0.314$_{-0.00917}^{+0.00904}$ &0.828$_{-0.0109}^{+0.0107}$ &0.96$_{-0.00694}^{+0.00699}$ &-1.04$_{-0.222}^{+0.223}$
&0.0996$_{-0.653}^{+0.657}$
&0.0487$_{-0.000616}^{+0.000612}$
&0.672$_{-0.011}^{+0.0107}$
&0.115  \\ 
CW&4&0.315$_{-0.0115}^{+0.0114}$ &0.828$_{-0.011}^{+0.0109}$ &0.961$_{-0.0072}^{+0.00728}$ &-1.03$_{-0.268}^{+0.263}$
&0.0669$_{-0.762}^{+0.769}$
&0.0487$_{-0.000597}^{+0.000608}$
&0.673$_{-0.0112}^{+0.0109}$
&0.1  \\ 
&&&&&&&&&\\ 
CX&0&0.317$_{-0.00881}^{+0.00881}$ &0.81$_{-0.0102}^{+0.00997}$ &0.968$_{-0.00698}^{+0.00686}$ &-1.36$_{-0.224}^{+0.234}$
&1.25$_{-0.887}^{+0.86}$
&0.0487$_{-0.000621}^{+0.000625}$
&0.674$_{-0.0115}^{+0.0115}$
&11.8  \\ 
CX&3&0.312$_{-0.00829}^{+0.00836}$ &0.826$_{-0.0111}^{+0.011}$ &0.96$_{-0.00717}^{+0.0072}$ &-1.07$_{-0.209}^{+0.207}$
&0.0629$_{-0.617}^{+0.62}$
&0.0487$_{-0.000599}^{+0.000611}$
&0.673$_{-0.0109}^{+0.0108}$
&0.324  \\ 
CX&4&0.313$_{-0.0108}^{+0.0108}$ &0.827$_{-0.0112}^{+0.0108}$ &0.961$_{-0.00699}^{+0.00682}$ &-1.05$_{-0.261}^{+0.26}$
&0.0731$_{-0.727}^{+0.731}$
&0.0486$_{-0.000598}^{+0.000605}$
&0.673$_{-0.0109}^{+0.0108}$
&0.478  \\ 
\hline
\end{tabular}
\end{center}
\label{tab:paradist_DESpp}
\end{table*}
 
 \begin{figure*}
\includegraphics[width=18cm]{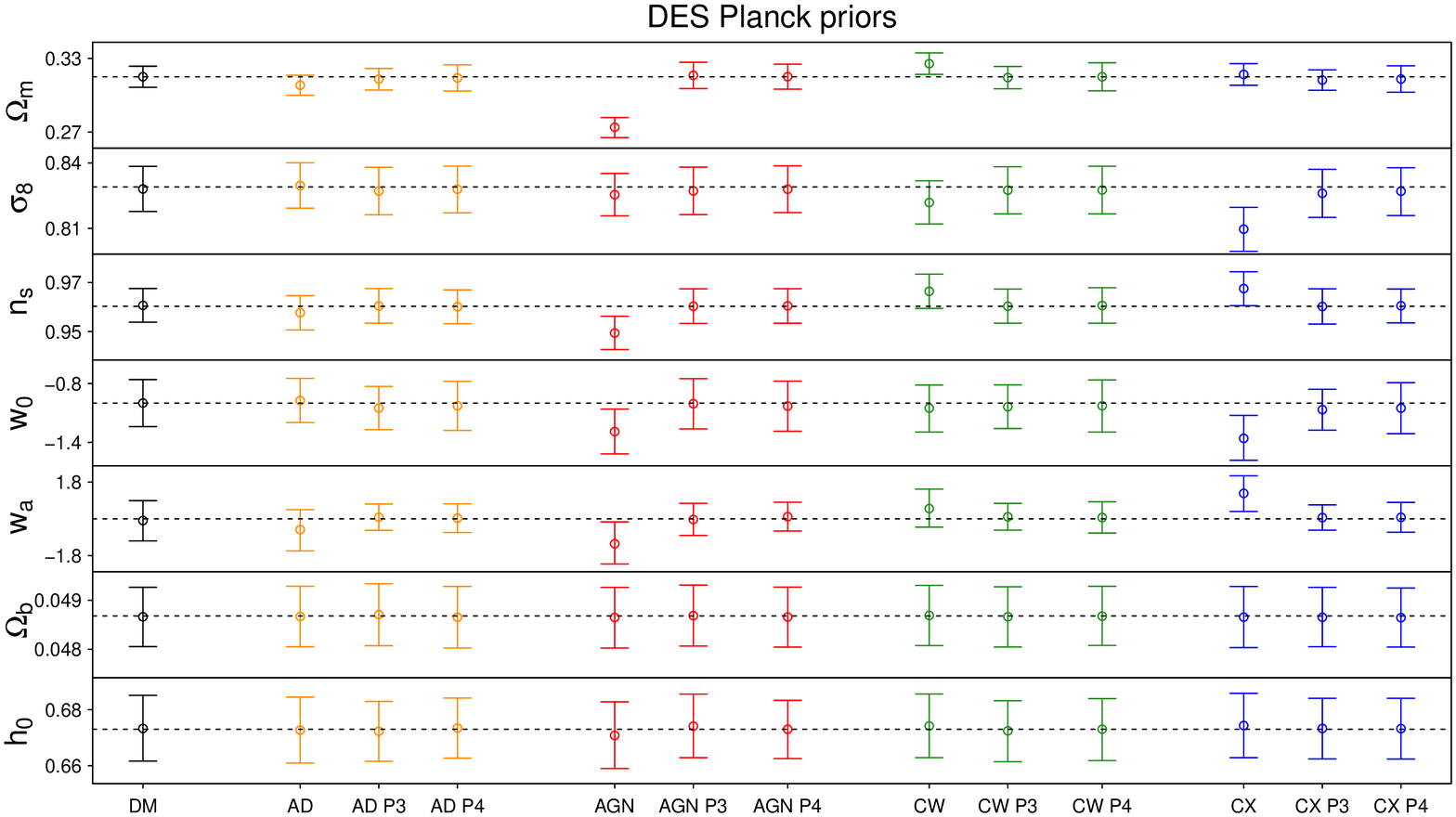}
\caption{The best fit and marginalized 1D error bars on cosmological parameters for an DES survey with Planck priors (see Table \ref{tab:paradist_DESpp} for exact numbers). }
         \label{fi:errorbias_despp}
\end{figure*}

\newpage 
 
\begin{figure*}
  \begin{minipage}[c]{0.85\textwidth}
\includegraphics[width=\textwidth]{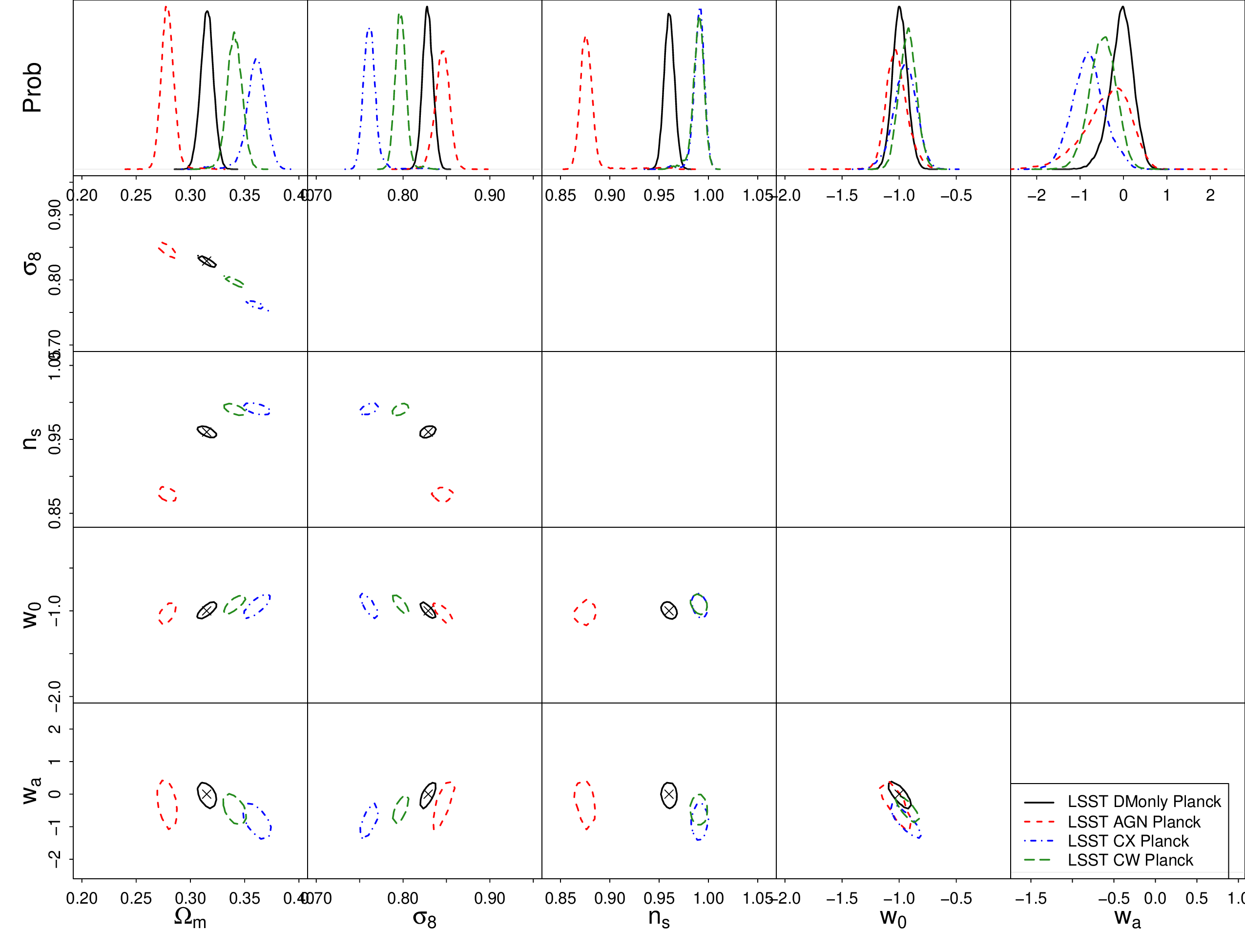}
  \end{minipage}\hfill
  \begin{minipage}[c]{0.13\textwidth}
   \caption{Cosmological constraints for a LSST/Euclid survey assuming different underlying baryonic scenarios for our Universe, i.e. pure dark matter \ti{(black, solid)}, strong AGN feedback \ti{(red, dashed)}, extreme cooling \ti{(blue, dot-dashed)}, and moderate cooling \ti{(green, long-dashed)}. The scenarios are detailed in Sect. \ref{sec:sim}. Results shown here include priors front the Planck mission.} \label{fi:lsst_pp}
  \end{minipage}
  \begin{minipage}[c]{0.85\textwidth}
\includegraphics[width=\textwidth]{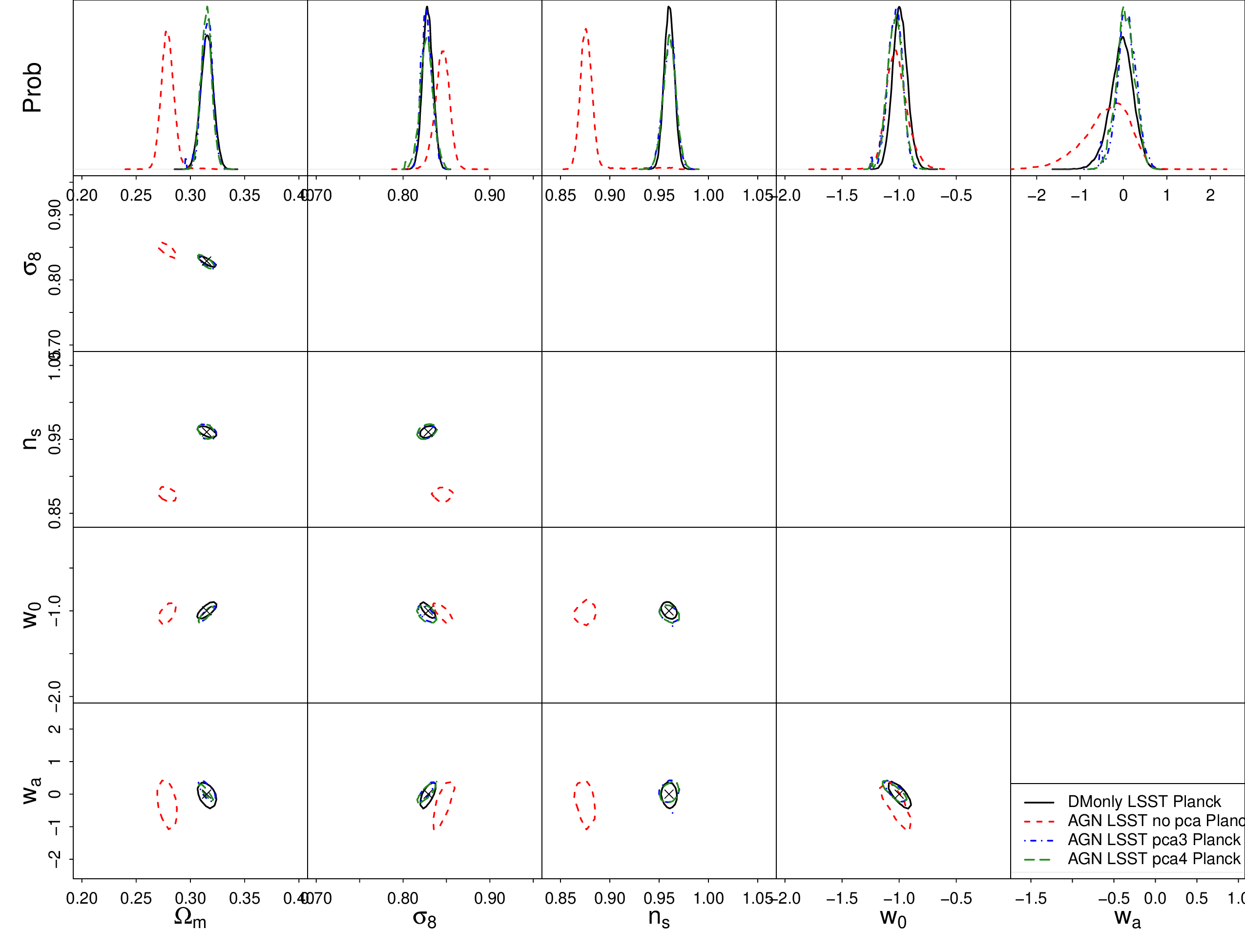}
  \end{minipage}\hfill
  \begin{minipage}[c]{0.13\textwidth}
      \caption{Cosmological constraints for a LSST/Euclid survey with Planck priors when using the PCA mitigation technique. The results shown assume that the baryonic physics of the Universe follows the AGN scenario (i.e. the most extreme baryonic scenario). We remove three and four PC modes (\ti{blue/dashed} and \ti{green/long-dashed}, respectively) and compare the results to the untreated AGN scenario (\ti{red/dashed}) and to a pure DM scenario (\ti{black/solid}).}\label{fi:lsst_pca_pp}
  \end{minipage}
\end{figure*}

\vspace{1cm}

\begin{table*}
\caption{Marginalized 1D constraints on cosmological parameters for the DM, AD, AGN, CW, and CX scenario with and without the PCA mitigation for a LSST/Euclid survey (with Planck priors). The last column contains the $\Delta \chi^2$ distance (see Eq. \ref{eq:paradistance}) between best fit and fiducial parameter point.}
\begin{center}
\def\arraystretch{1.3}
\begin{tabular}{|c c c c c c c c c c|}
\hline
Scenario & PCA order & $\om$ & $\sig$ & $\ns$ & $\w$ &$\wa$ & $\omb$ & $h_0$ & $\Delta \chi^2_\mr{BA}$ \\
\hline
DM&0&0.315$_{-0.00584}^{+0.00577}$ &0.828$_{-0.00591}^{+0.00604}$ &0.96$_{-0.00513}^{+0.00519}$ &-0.993$_{-0.0635}^{+0.0638}$
&-0.0449$_{-0.258}^{+0.26}$
&0.0486$_{-0.000604}^{+0.000608}$
&0.673$_{-0.0098}^{+0.0096}$
&0.096  \\ 
&&&&&&&&&\\ 
AD&0&0.301$_{-0.00569}^{+0.00561}$ &0.846$_{-0.00629}^{+0.00643}$ &0.944$_{-0.00539}^{+0.0053}$ &-1.04$_{-0.0607}^{+0.0604}$
&0.245$_{-0.259}^{+0.258}$
&0.0485$_{-0.000591}^{+0.00061}$
&0.683$_{-0.0103}^{+0.0103}$
&60.7  \\ 
AD&3&0.315$_{-0.00506}^{+0.00506}$ &0.826$_{-0.00638}^{+0.0064}$ &0.96$_{-0.00645}^{+0.00623}$ &-0.997$_{-0.0622}^{+0.0608}$
&-0.111$_{-0.224}^{+0.216}$
&0.0487$_{-0.000621}^{+0.000624}$
&0.671$_{-0.00924}^{+0.00911}$
&2.66  \\ 
AD&4&0.315$_{-0.00482}^{+0.00474}$ &0.828$_{-0.00668}^{+0.00712}$ &0.96$_{-0.00656}^{+0.00636}$ &-1$_{-0.0591}^{+0.0651}$
&-0.0452$_{-0.202}^{+0.204}$
&0.0486$_{-6e-04}^{+0.000624}$
&0.673$_{-0.00932}^{+0.00977}$
&0.643  \\ 
&&&&&&&&&\\ 
AGN&0&0.279$_{-0.00585}^{+0.00559}$ &0.845$_{-0.00812}^{+0.00808}$ &0.878$_{-0.00775}^{+0.00433}$ &-1.02$_{-0.0879}^{+0.0905}$
&-0.4$_{-0.53}^{+0.517}$
&0.0478$_{-0.000624}^{+0.000686}$
&0.702$_{-0.00988}^{+0.0128}$
&207  \\ 
AGN&3&0.315$_{-0.0051}^{+0.00517}$ &0.827$_{-0.00635}^{+0.00626}$ &0.96$_{-0.00646}^{+0.00619}$ &-1.04$_{-0.0625}^{+0.0631}$
&0.0715$_{-0.206}^{+0.222}$
&0.0487$_{-0.000613}^{+0.000623}$
&0.674$_{-0.00899}^{+0.00891}$
&2.85  \\ 
AGN&4&0.315$_{-0.00505}^{+0.00479}$ &0.827$_{-0.0072}^{+0.00745}$ &0.961$_{-0.00656}^{+0.00638}$ &-1.04$_{-0.0663}^{+0.0658}$
&0.0446$_{-0.207}^{+0.216}$
&0.0487$_{-0.000623}^{+0.000605}$
&0.674$_{-0.00938}^{+0.00937}$
&6.6  \\ 
&&&&&&&&&\\ 
CW&0&0.341$_{-0.00671}^{+0.00694}$ &0.797$_{-0.00629}^{+0.00594}$ &0.989$_{-0.00512}^{+0.00577}$ &-0.924$_{-0.0757}^{+0.074}$
&-0.477$_{-0.298}^{+0.304}$
&0.049$_{-0.000633}^{+0.000634}$
&0.652$_{-0.0113}^{+0.0107}$
&75.1  \\ 
CW&3&0.315$_{-0.00469}^{+0.00477}$ &0.829$_{-0.00631}^{+0.00629}$ &0.96$_{-0.00656}^{+0.00655}$ &-0.995$_{-0.0561}^{+0.0559}$
&9.19e-05$_{-0.202}^{+0.203}$
&0.0487$_{-0.000604}^{+0.000608}$
&0.672$_{-0.00881}^{+0.00885}$
&1.64  \\ 
CW&4&0.315$_{-0.00477}^{+0.00477}$ &0.829$_{-0.00717}^{+0.00728}$ &0.96$_{-0.0064}^{+0.00643}$ &-0.994$_{-0.0573}^{+0.0564}$
&-0.0112$_{-0.199}^{+0.203}$
&0.0487$_{-0.000604}^{+0.00061}$
&0.672$_{-0.00942}^{+0.00927}$
&0.244  \\ 
&&&&&&&&&\\ 
CX&0&0.36$_{-0.00793}^{+0.00885}$ &0.762$_{-0.00731}^{+0.00607}$ &0.99$_{-0.00481}^{+0.00549}$ &-0.95$_{-0.0982}^{+0.0995}$
&-0.806$_{-0.368}^{+0.368}$
&0.0492$_{-0.000625}^{+0.000647}$
&0.624$_{-0.0125}^{+0.0107}$
&154  \\ 
CX&3&0.315$_{-0.00476}^{+0.00486}$ &0.824$_{-0.00612}^{+0.00605}$ &0.96$_{-0.00663}^{+0.00664}$ &-1.01$_{-0.0601}^{+0.0599}$
&-0.148$_{-0.22}^{+0.215}$
&0.0487$_{-0.000601}^{+0.000609}$
&0.672$_{-0.0091}^{+0.00896}$
&2.22  \\ 
CX&4&0.316$_{-0.00507}^{+0.00507}$ &0.825$_{-0.00687}^{+0.00672}$ &0.959$_{-0.00681}^{+0.0068}$ &-1.01$_{-0.0639}^{+0.0635}$
&-0.0895$_{-0.213}^{+0.21}$
&0.0487$_{-0.000605}^{+0.000596}$
&0.673$_{-0.00931}^{+0.00947}$
&6.82  \\ 
\hline
\end{tabular}
\end{center}
\label{tab:paradist_LSSTpp}
\end{table*}

\begin{figure*}
\includegraphics[width=18cm]{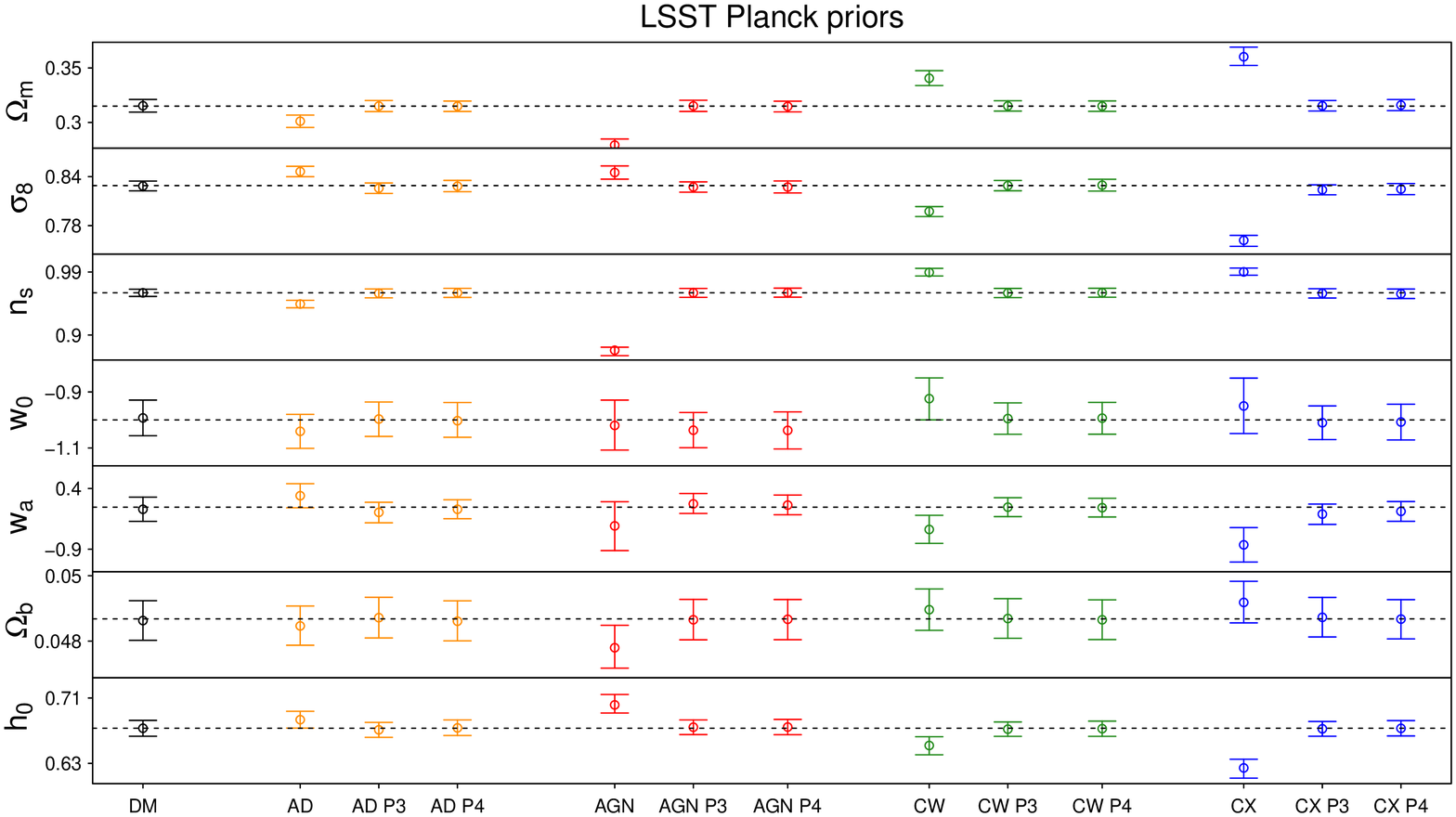}
\caption{The best fit value and marginalized 1D error bars on cosmological parameters for an LSST/Euclid  survey with Planck priors (see Tables \ref{tab:paradist_LSSTpp} for exact numbers). }
         \label{fi:errorbias_lsstpp}
\end{figure*}

\end{appendix}

\end{document}

%% file: journaldef.tex
\def\apj{ApJ}%
\def\apjl{ApJ}%
\def\apjs{ApJS}%
\def\aap{A\&A}%
\def\mnras{MNRAS}%
\def\prd{Phys.~Rev.~D}%
\def\pasp{PASP}%
\def\jcap{JCAP}%
\def\physrep{Phys.~Rep.}%